\documentclass[preprint, 10pt]{aastex}       

\newcommand{\der}[2][\;\;]{\ensuremath{ \frac{d{#1}}{d{#2}} }}
\newcommand{\dern}[3][\;\;]{\ensuremath{ \frac{d^{#3}{#1}}{d{#2}^{#3}} }}
\newcommand{\dpar}[2][\;\;]{\ensuremath{ \frac{\partial{#1}}{\partial{#2}} }}

\newcommand{\bvec}[1]{{\mbox{{\boldmath$#1$}}}}		
\newcommand{\unitv}[1]{\bvec{\hat{#1}}}			
\newcommand{\grad}{\bvec{\nabla}}			

\newcommand{\Hc}{{\cal H}}				
\newcommand{\dP}{\delta\varpi}				

\newcommand{\GGG}{GGG}	 	
\newcommand{\gafd}{GApFD}	
\newcommand{\jfm}{JFM}		
\newcommand{\jpo}{JPO}		
\newcommand{\natas}{NatAs}	
\newcommand{\ptrslA}{RSPTA}	
\newcommand{\rsptA}{RSPTA}	
\newcommand{\phfl}{PhFl}	
\newcommand{\qjmam}{QJMAM}      
\newcommand{\sci}{Sci}		

\newcommand{\eqnref}[1]{(\ref{#1})}

\begin{document}

\title{Radial Trapping of Thermal Rossby Waves within the Convection Zones of Low-Mass Stars}
\shorttitle{Thermal Rossby Waves within Low-Mass Stars}

\author{Bradley W. Hindman}
\affil{JILA, University of Colorado, Boulder, CO~80309-0440, USA}
\affil{Department of Applied Mathematics, University of Colorado, Boulder, CO~80309-0526, USA}
\email{hindman@solarz.colorado.edu}

\author{Rekha Jain}
\affil{School of Mathematics \& Statistics, University of Sheffield, Sheffield S3 7RH, UK}


\begin{abstract}

We explore how thermal Rossby waves propagate within the gravitationally stratified
atmosphere of a low-mass star with an outer convective envelope. Under the conditions
of slow, rotationally constrained dynamics, we derive a local dispersion relation
for atmospheric waves in a fully compressible stratified fluid. This dispersion
relation describes the zonal and radial propagation of acoustic waves and gravito-inertial
waves. Thermal Rossby waves are just one class of prograde-propagating gravito-inertial
wave that manifests when the buoyancy frequency is small compared to the rotation
rate of the star. From this dispersion relation, we identify the radii at which
waves naturally reflect and demonstrate how thermal Rossby waves can be trapped
radially in a waveguide that permits free propagation in the longitudinal direction.
We explore this trapping further by presenting analytic solutions for thermal Rossby
waves within an isentropically stratified atmosphere that models a zone of efficient
convective heat transport. We find that within such an atmosphere, waves of short
zonal wavelength have a wave cavity that is radially thin and confined within the
outer reaches of the convection zone near the star's equator. The same behavior is
evinced by the thermal Rossby waves that appear at convective onset in numerical
simulations of convection within rotating spheres. Finally, we suggest that stable
thermal Rossby waves could exist in the lower portion of the Sun's convection zone,
despite that region's unstable stratification. For long wavelengths, the Sun's rotation
rate is sufficiently rapid to stabilize convective motions and the resulting overstable
convective modes are identical to thermal Rossby waves.

\end{abstract}



\section{Introduction}
\label{sec:Introduction}

The unambiguous detection of Rossby waves in the Sun by \cite{Loeptien:2018} has
led to a flurry of observational efforts to characterize the waves and to search
for other classes of inertial oscillations \citep[e.g.,][]{Liang:2019, Hanasoge:2019,
Alshehhi:2019, Hanson:2020, Proxauf:2020, Gizon:2021, Hathaway:2021, Mandal:2021}.
As a result, there has been a resurrection of interest in inertial waves as they
might apply to the Sun and solar-like stars \citep[e.g.,][]{Lanza:2019, Damiani:2020,
Gizon:2020, Cai:2021, Bekki:2022}. In particular, \cite{Gizon:2021} have raised the
intriguing possibility that inertial oscillations might be a sensitive seismic diagnostic
for the properties of the deep convection zone, such as the radial profiles of the
superadiabatic gradient and the turbulent viscosity. For a recent review of inertial
oscillations in the solar context, see the review by \cite{Zaqarashvili:2021}.

There is a menagerie of inertial oscillations that might exist within a star
\citep[e.g.,][]{Bryan:1889, Greenspan:1968, Rieutord:1991, Lindblom:1999, Lockitch:1999}.
But, the classical Rossby waves---coined $r$ modes in astrophysics
\citep{Papaloizou:1978}---have received the bulk of the attention of the stellar
physics community \citep[e.g.,][]{Provost:1981, Saio:1982, Wolff:1986, Lee:1987,
Lee:1992,  Saio:2018, VanReeth:2018, Li:2019}. The $r$ modes are primarily 2D
vorticity waves, with incompressive horizontal motions confined largely to spherical
surfaces. As such motions do not generate pressure or buoyancy fluctuations, the
Coriolis force is the only active restoring force.

Classical Rossby waves conserve the radial component of their absolute vorticity.
If $\bvec{\Omega}$ is the rotation vector of the star and the flow velocity in
the rotating frame of reference is $\bvec{u}$, then the relative vorticity of the
flow is $\bvec{\omega} = \bvec{\nabla} \times \bvec{u}$ and the ``planetary" vorticity
is $2\bvec{\Omega}$. The absolute vorticity is their sum,
$\bvec{\omega}_{a} = \bvec{\omega} + 2\bvec{\Omega}$, and the conservation principle
can be written, 

\begin{equation}
	\frac{D}{Dt}\left(\bvec{\omega}_a\cdot \unitv{r}\right)
		= \frac{D}{Dt}\left(\omega_r + 2\Omega_r\right) = 0 \; ,
\end{equation}

\noindent where $\omega_r$ and $\Omega_r = \Omega \, \cos\theta$ are the radial
components of the relative vorticity and the rotation vector, respectively. The
angle $\theta$ is the colatitude and $\unitv{r}$ is the radial unit vector of the
spherical coordinate system whose axis is aligned with the rotation vector.

In the northern hemisphere, as a spinning parcel of fluid is pushed northward, the
radial component of the planetary vorticity $2\Omega_r$ increases and there must
be a corresponding decrease in the relative vorticity $\omega_r$. A parcel pushed
equatorward has the converse effect; the radial component of the planetary vorticity
is reduced and the relative vorticity must increase to compensate. For the poleward
pushed parcel, the decrease in relative vorticity is consistent with adding an
anticyclonic vortex to the fluid parcel and through vortex-vortex interactions this
new clockwise motion will push all nearby fluid elements. Fluid located to the east
is pushed poleward and fluid to the west is pushed equatorward. These pushed fluid
elements will conserve their own absolute vorticity and in turn push fluid elements
that are further and further away. One can see that the net result is a north-south
undulation that travels to the west, or equivalently in the {\bf retrograde} direction.
In the preceding discussion, we are using the words east and west in terms of the
directions on the stellar globe, {\it not} in terms of the directions on the plane
of the sky (as is often done by solar astronomers).

There is another class of Rossby wave, called the thermal Rossby wave
\citep[i.e.,][]{Roberts:1968, Busse:1970}, that operates on a similar principle
of conservation of potential vorticity. The motions in a thermal Rossby wave are
also 2D, but instead of motion on spherical surfaces, the thermal Rossby wave has
motions that are perpendicular to the rotation axis. Specifically, the motion
generally manifests as a belt of nearly geostrophic Taylor columns that gird the
equator and, except for where they intersect with the star's spherical surface,
are largely invariant in the direction of the rotation vector---see the review by
\cite{Busse:2002}. Every other column in the belt has right-handed spin and columns
with left-handed spin are interleaved between. A full longitudinal wavelength is
hence two counter-spinning columns.

Each spinning Taylor column conserves angular momentum locally and when expressed
as a potential vorticity, the conservation law takes on the form \citep{Gibbons:1980,
Glatzmaier:1981c, Unno:1989},

\begin{equation}
	\frac{D}{Dt} \left( \frac{\bvec{\omega}_a\cdot\unitv{\Omega}}{\rho L}\right)
		= \frac{D}{Dt} \left( \frac{\omega_y + 2\Omega}{\rho L}\right) = 0 \; ,
	\label{eqn:potential_vorticity}
\end{equation}

\noindent where $\omega_y = \bvec{\omega}\cdot\unitv{\Omega}$ is the component of
the relative vorticity aligned with the rotation axis, $\rho$ is the mass density,
and $L$ is the height of the Taylor column, i.e., the length of the chord that runs
along the column's axis from the stellar surface in the southern hemisphere to the
same spherical surface in the north---see Figure 1 of \cite{Glatzmaier:1981c}.

If a spinning column near the equator is pushed towards the rotation axis, the column
grows in height ($L$ increases) as the chord length of the column's axis increases.
In an incompressible fluid, this vortex stretching is accompanied by a commensurate
narrowing of the column to conserve mass. Subsequently, as the column compresses
laterally the column must spin faster to conserve angular momentum about its own axis
\citep[e.g.,][]{Hide:1966, Busse:1970}. In Equation~\eqnref{eqn:potential_vorticity},
this conservation principle is enforced by the constancy of the potential vorticity,
$(\omega_y+2\Omega) / \rho L$. As $L$ increases, $\omega_y$ must also
increase. The resulting induced vorticity causes the neighboring column
to the west to be pushed outward, away from the rotation axis and the column to the
east to be pushed inwards towards the rotation axis. These newly pushed columns conserve
their own potential vorticity (i.e., angular momentum) and induce spinning columns further
down the belt to also move inward and outward. The result is a {\bf prograde} propagating
Rossby wave where the spinning columns dance back and forth, toward and away from the
rotation axis. This type of wave has received significant attention in the geophysics
community, as {\it thermal} convection in a rotating spherical shell of fluid appears
at onset as an unstable {\it thermal} Rossby wave \citep[e.g.,][]{Roberts:1968, Busse:1970,
Dormy:2004, Jones:2009, Kaplan:2017}.

In a compressible fluid, an additional mechanism comes into play \citep{Gibbons:1980,
Glatzmaier:1981c, Ando:1989}. In complement to the topological effect of the changing
column height as the column moves inward or outward, gravitational stratification
leads to a change in the mass density of the column as the column is pushed into
a region with different pressure. If the column is pushed closer to the rotation axis,
the increase in pressure leads to an increase in density and conservation of mass and
angular momentum dictate that the column must narrow and consequently spin faster
cyclonically. This stratification effect is embodied by the factor
of density that appears in the denominator of the potential vorticity in
Equation~\eqnref{eqn:potential_vorticity}.

The topological and stratification effects act in concert to enhance spin as a column
is moved inward, and hence they both lead to prograde wave propagation. The topological
effect, sometimes called the topological $\beta$-effect, is the dominant effect in the
molten interiors of rocky planets, as the density contrast as a function of radius is
rather modest and one can safely assume that the fluid is incompressible. In stars, however,
many density scale heights fit within the stellar radius and as a result, the topological
effect is usually ancillary to the stratification. For instance, \cite{Glatzmaier:1981c}
argue that the stratification effect dominates if a star spans more than one density scale
height (the Sun's convection zone spans roughly eleven scale heights).

To our knowledge, thermal Rossby waves have not yet been observed in the Sun or
identified asteroseismically in other stars. Yet, they are a conspicuous feature in
laboratory experiments of convection in a rotating fluid \citep[e.g.,][]{Mason:1975,
Busse:1982,  Azouni:1985, Chamberlain:1986, Sommeria:1991, Cordero:1992, Smith:2014}.
In fact, thermal Rossby waves are often used as a proxy for understanding classical
Rossby waves, by building a spinning tank of fluid with sloping upper or lower boundaries,
thus allowing one to control the sign and magnitude of the topological $\beta$-effect.
Further, in numerical simulations of stellar convection in spherical geometry, nonlinear
thermal Rossby waves play a crucial role in the transport of angular momentum and heat
throughout the star's convection zone. In particular, they are essential for generating
differential rotation by redistributing angular momentum. These waves appear prominently
at convective onset, when the Rayleigh number just exceeds the critical value required
for convective overturning. The first convective instability to appear is a buoyantly
driven thermal Rossby wave that manifests at low latitudes just below the outer surface.
In terms of spherical harmonics, this wave is composed of a single unstable azimuthal
order $m$ and a small range of harmonic degrees corresponding to equatorially trapped
harmonics, $\ell \approx m$. The wave is unstable, but its amplitude saturates at a low
amplitude due to nonlinear effects. Figure~\ref{fig:thermal_Rossby} illustrates the type
of thermal Rossby wave that is seen in such simulations \citep[i.e.,][]{Hindman:2020a}.
Both panels show the axial component of the vorticity (i.e., the component that is parallel
to the rotation vector, $\omega_y = \bvec{\omega}\cdot\unitv{\Omega}$). The rotation
axis is aligned with the coordinate axis, with north upwards in the image. The left-hand
panel is an orthographic projection of the axial vorticity on a spherical surface with
a radius that is just below the outer surface of the simulation domain. Red tones indicate
anticyclonic vorticity and blue tones cyclonic. This particular simulation has 42 complete
wavelengths that wrap around the equator ($m = 42$). The right-hand panel of
Figure~\ref{fig:thermal_Rossby} portrays axial vorticity in the equatorial plane.
It is clear that each zonal wavelength consists of two counter-rotating vortices and
that the wave is trapped in the upper reaches of the model's convection zone.

This concentration of the thermal Rossby wave's energy density in the outer portion
of the star's convection zone is a direct consequence of the density stratification.
In an incompressible fluid, the thermal Rossby waves that appear at convective onset
are concentrated in the deepest portions of the shell. But with even a moderate degree
of stratification---one or more density scale heights across the shell, the wave's
energy density begins to shift and cling to the outer surface \citep[e.g.,][]{Busse:2005,
Jones:2009,Hindman:2020a}. The radial structure of a thermal Rossby wave's eigenfunction
and the way in which the waves are reflected and trapped in a waveguide are still
poorly understood when the fluid is stratified. Elucidating why these reflections occur 
and where the waveguide resides will be our goal here.

In helio- and asteroseismology, when exploring the properties of the waveguides that trap
$p$ modes and $g$ modes, it is common practice to extract a local dispersion relation from
the fluid equations and to discern the regions of a star where a wave with a particular
frequency and horizontal wavenumber is radially propagating and where it is evanescent
\citep[e.g.,][]{Unno:1989, Gough:1993, Christensen-Dalsgaard:2003}. In this way, the radial
extent of the waveguide can be identified as the zone of propagation without ever needing
to completely solve for the mode eigenfunctions.  We will apply this standard technique
here to define the radial boundaries of the waveguide for thermal Rossby waves and to
characterize which properties of the star (e.g., the stratification and rotation rate)
determine the location of those boundaries.

A local dispersion relation for thermal Rossby waves has been derived previously by
\cite{Ando:1989} and subsequently simplified by \cite{Unno:1989} for the special case
of waves propagating perpendicular to the rotation vector through an isentropic stratification.
If $\omega$ is the temporal frequency of the wave (not to be confused with the vorticity),
$k_x$ is the zonal wavenumber, $k_z$ is the radial wavenumber, and $H$ is the density
scale height, \cite{Unno:1989} obtain the following relation 

\begin{equation}
	\omega = \frac{2\Omega k_x}{\left(k_x^2+k_z^2\right) H} \; .
	\label{eqn:Unno_dispersion}
\end{equation}

\noindent Both, \cite{Ando:1989} and \cite{Unno:1989} correctly comment that the wave
propagates in the prograde direction and, as such, \cite{Unno:1989} calls the thermal
Rossby wave the ``low-frequency prograde wave." As we will see, their expression is
valid only in the extreme limit that the wavelength is drastically shorter than the
local density scale height. Thus, in their derivation they have dropped a variety of
terms that depend on the star's radial stratification. We will demonstrate that these
dropped terms can cause reflection and thus enable radial trapping of the thermal
Rossby wave.

In addition to deriving a more complete local dispersion relation that self-consistently
accounts for the gravitational stratification, we will explore the nature of the
resulting waveguide, or wave cavity, by presenting analytic solutions for the
eigenfunctions and eigenfrequencies that apply when the atmosphere is isentropically
stratified. Such stratification has been examined often in the past
\citep[e.g.,][]{Glatzmaier:1981c, Papaloizou:1981, Busse:2005, Wu:2005, Busse:2014,
Ouazzani:2020} because in such an atmosphere the internal gravity waves and their
coupling to the inertial waves is suppressed. With these eigenfunctions we will
demonstrate that when the thermal Rossby wave is a pure inertial wave (instead of
a mixed gravito-inertial wave), the concentration of the thermal Rossby wave's 
energy density near the upper boundary of a stellar convection zone is a natural
consequence for waves with a zonal wavelength that is much shorter than the depth of
the convection zone.

In Section~\ref{sec:Atmospheric_Waves}, we present the fluid equations for a
completely compressible fluid and argue that the solutions are 2D when the wave
motions are sufficiently slow and the frequency is low. From these equations, in
Section~\ref{sec:Local_Dispersion}, we derive a local dispersion relation for
gravito-inertial waves that is valid for a completely general radial stratification.
In Section~\ref{sec:Neutrally_Stable}, we specialize to an isentropically stratified
atmosphere and present analytic solutions for the resulting inertial waves. Finally,
in Section~\ref{sec:Discussion}, we discuss the implications of our results for
the Sun and other low-mass stars.


\section{Atmospheric Waves in a Rotating Star}
\label{sec:Atmospheric_Waves}

Our ultimate goal is to generate a simple, easily-interpreted local dispersion
relation for thermal Rossby waves that is applicable in a gravitationally stratified
atmosphere. From this dispersion relation we will deduce where waves reflect in the
atmosphere and thus map the radial extent of the waveguide or cavity in which the
thermal Rossby waves are trapped. In this section we will derive the governing
wave equation that applies in a simplified geometry. 

We will be implicitly thinking about how the waves propagate and reflect within
a low-mass star that possesses an outer convection zone, but many of the ideas
that we develop will apply to more massive stars as well. In a general stratification,
atmospheric waves come in many forms: acoustic waves, internal gravity waves, surface
gravity waves, inertial waves, and all of their potential hybrids. In a star, the
acoustic waves, or $p$ modes, generally appear at high frequency and thus their
behavior can often be separated from that of the coupled, low-frequency, gravito-inertial
waves. Thermal Rossby waves are but one type of the general class of gravito-inertial
waves.

\subsection{2D Rotationally Constrained Flow in a Local Cartesian Domain}
\label{subsec:2D_rotational_constraint}

In regions of a star that are stably stratified, vertical movement is strongly
inhibited and motions are often assumed to be horizontal and confined to spherical
surfaces. When deriving the $r$ modes of a star, this assumption is called the
traditional approximation of rotation. In regions with an unstable or neutrally
stable stratification, vertical motion is no longer restricted and the traditional
approximation is not appropriate for all inertial waves. Instead, we start by
assuming that the motions are rotationally constrained (i.e., slow compared to the
rotational speeds) and, hence, the Taylor-Proudman theorem applies. This theorem
dictates that slow motions are invariant in the direction of the rotation vector
$\bvec{\Omega}$ and, therefore, we can treat the problem as two-dimensional (2D).
This 2D approximation is a standard one \citep{Busse:1970, Busse:2002, Busse:2005,
Busse:2014, Glatzmaier:1981c}, but of course, in a star with spherical geometry,
there are many reasons that the Taylor-Proudman theorem breaks down on global
spatial scales. Even so, the Coriolis force leads to a strong tendency towards
2D motion, and the Taylor-Proudman theorem can still hold locally over surprisingly
long length scales. In our work here, we will ignore these potential 3D effects,
because the full spherical geometry begets complicated mathematics that hinders
qualitative understanding.

Numerical simulations \citep[e.g.,][]{Hindman:2020a} and linear stability analyses
\citep{Soward:1983, Yano:1992, Jones:2009} demonstrate that the thermal Rossby waves
that appear at convective onset in a gravitationally stratified spherical shell have
the properties that the waves propagate in the longitudinal direction and are essentially
2D in nature. The waves appear as a parade of convective rolls that wrap around the
equator. The axis of each roll is aligned with the rotation vector and the roll is
largely invariant along its own axis except where the axis intersects with the outer
spherical surface. Whenever the zonal wavelength is much smaller than the outer radius
of the convection zone, the waves are confined within a radially thin cavity that hugs
the outer surface of the convection zone near the equator. Further,
since the waves are essentially 2D in nature, the waves are equatorially as well as
radially trapped. For example, the bottom of the wave cavity at the equator can be
projected along a cylindrical surface whose axis is perpendicular to the equatorial
plane. Hence, the cavity is confined in cylindrical radius instead of spherical radius
and modes with a deep cavity (at the equator) extend further north and south of the
equator. Figure~\ref{fig:thermal_Rossby} illustrates these properties by presenting
a thermal Rossby wave as it appears in the 3D numerical simulation of \cite{Hindman:2020a}. 
Other examples can be viewed and downloaded at \cite{Hindman:2020b}. The 
numerically computed eigenfunctions obtained by \cite{Bekki:2022} in spherical geometry
also illustrate the cylindrical nature of the lower boundary of the wave cavity.

We will exploit these properties and make a short longitudinal wavelength approximation.
Thus, over the shallow layer within which the wave resides, we can ignore the radial
variation in the gravitational acceleration. Further, we can ignore the curvature
of the isopycnals. These assumptions allow us to treat the background fluid as a
plane-parallel atmosphere with constant gravity. This is similar to the approximation
that is  often made for the $p$ modes in helioseismology \citep[e.g.,][]{Lamb:1945,
Gough:1993, Hindman:1994}; for large harmonic degrees (i.e., large horizontal wavenumbers),
the acoustic waves are trapped in a thin planar waveguide that lies just below the
photosphere.

We will solve the fluid equations in a local Cartesian coordinate system whose
origin is located at the equator on the star's outer surface. We align the unit
vectors, $\unitv{x}$, $\unitv{y}$, and $\unitv{z}$, of this coordinate system
such that $\unitv{x}$ points in the longitudinal direction, $\unitv{y}$ points in
the latitudinal direction (parallel to the rotation vector, $\bvec{\Omega} = \Omega \unitv{y}$),
and $\unitv{z}$ is aligned with the radial direction (or antialigned with gravity,
$\bvec{g}=-g\unitv{z}$). If $r$, $\theta$, and $\phi$ are the radius, colatitude,
and longitude of a spherical coordinate system whose axis is aligned with the rotation
vector, the three Cartesian coordinates can be mapped onto the spherical coordinates
through $x = R \phi$, $y = R \left(\pi/2 - \theta\right)$, and $z = r - R$, where $R$
is the radius of the stellar photosphere. We will interchangeably refer to $z$ as the
radial or vertical coordinate, and $x$ and $y$ as horizontal coordinates. This Cartesian
coordinate system is similar to an $f$-plane model, except that the rotation vector
points horizontally in the latitudinal direction. In a traditional $f$-plane model the
horizontal component of the rotation vector is ignored and the {\it vertical} component
is treated as constant. Here, we will instead ignore the vertical component and impose
homogeneity of the {\it latitudinal} component.

\subsection{Background Atmosphere for a General Stratification}
\label{subsec:background}

As is typically done, we decompose all of the fluid variables into a steady background
term and a temporally oscillating fluctuation, where the background is denoted with
a `0' subscript and the Eulerian fluctuations with `1'. Since the gravity in our
plane-parallel atmosphere is purely vertical, all of the thermodynamic variables of
the background are functions of $z$ alone. Hence, the background pressure, mass density,
and temperature are, respectively, $P_0(z)$, $\rho_0(z)$, and $T_0(z)$. These background
thermodynamic profiles obey the hydrostatic condition and the ideal gas law,

\begin{eqnarray}
	\der[P_0]{z} = - g \rho_0 \; ,
	\label{eqn:hydrostatic}
\\
	P_0 = \rho_0 R_{\rm gas} T_0 \; ,
\end{eqnarray}

\noindent where $R_{\rm gas}$ is the gas constant.
	
Since we will be considering gravito-inertial waves, the buoyancy frequency, $N$,
will be an important atmospheric profile. The buoyancy frequency can be expressed
in terms of the density scale height, $H$, and the sound speed, $c$,

\begin{eqnarray}
	N^2 &=& g\left(\frac{1}{H} - \frac{g}{c^2}\right) \; ,
	\label{eqn:def_N}
\\
	H^{-1} &\equiv& -\frac{1}{\rho_0} \der[\rho_0]{z} \; ,
\\
	c^2 &\equiv&  \gamma R_{\rm gas} T_0 \; ,
\end{eqnarray}

\noindent where $\gamma$ is the gas's adiabatic exponent. Throughout
Sections~\ref{sec:Atmospheric_Waves} and \ref{sec:Local_Dispersion}, we shall
consider a general stratification where any reasonable functional form for the
atmospheric profiles will be possible. Later, in Section~\ref{sec:Neutrally_Stable},
we will specialize to an isentropic atmosphere for which $N^2 = 0$.

\subsection{Wave Equation for Atmospheric Waves}
\label{subsec:wave_equation} 

In our plane-parallel atmosphere, the 2D nature of the gravito-inertial waves manifests
as invariance along the $y$ axis and a lack of motion in that same direction. Thus,
the fluid velocity $\bvec{u}$ is purely lateral to the rotation vector $\bvec{\Omega}$
and depends only on time and on the longitudinal and radial coordinates, $x$ and $z$,
respectively, $\bvec{u} = u(x,z,t) \, \unitv{x} + w(x,z,t) \, \unitv{z}$. Further,
if we ignore ionization and diffusive effects, the linearized fluid equations for
an ideal gas take on the following form 

\begin{eqnarray}
	\dpar[\bvec{u}]{t} &=& 2  \bvec{u} \times \bvec{\Omega} - \frac{1}{\rho_0} \grad P_1
		+ \bvec{g} \frac{\rho_1}{\rho_0} \; ,
	\label{eqn:vector_momentum}
\\
	\dpar[P_1]{t} + w \der[P_0]{z} &=& c^2 \left(\dpar[\rho_1]{t} + w\der[\rho_0]{z}\right) \; ,
	 \label{eqn:adiabaticity}
\\	
	\dpar[\rho_1]{t} &=& -\grad \cdot \left(\rho_0\bvec{u}\right) \; ,
	\label{eqn:compressibility}
\end{eqnarray}

\noindent where we have adopted an adiabatic energy equation~\eqnref{eqn:adiabaticity}
and a continuity equation~\eqnref{eqn:compressibility} that describes a completely
compressible fluid. The quantities $\rho_1$ and $P_1$ are the Eulerian fluctuations
of the mass density and gas pressure, respectively.

Since the background atmosphere is steady and longitudinally invariant, we will
assume plane-wave solutions for the longitudinal and temporal variables. Therefore,
all fluid variables will have the following form

\begin{equation} \label{eqn:Fourier_Convention}
	w \propto e^{i k_x x}  \, e^{-i\omega t} \; ,
\end{equation}

\noindent with $k_x$ being the longitudinal wavenumber and $\omega$ the temporal
frequency. When a correspondence is drawn with spherical coordinates, the zonal
wavenumber is directly proportional to the azimuthal order $m = k_x R$ of the
associated spherical harmonic. Note, the sign convention that we have adopted is
such that waves with positive wavenumbers ($k_x>0$) have a prograde phase speed
if the frequency is positive ($\omega>0$) and retrograde for negative frequencies
($\omega<0$).

Our experience with acoustic-gravity waves suggests that a clean working variable
is the reduced Lagrangian pressure fluctuation, $\dP$, which is related to the
Eulerian pressure fluctuation through an additive advective term,

\begin{equation} \label{eqn:Lagrangian_pressure}
	\dP \equiv \frac{P_1}{\rho_0} - \frac{\bvec{u} \cdot \grad P_0}{i\omega\rho_0} = \frac{P_1}{\rho_0} + \frac{gw}{i\omega} \; .
\end{equation}

\noindent By utilizing the definition of the buoyancy frequency \eqnref{eqn:def_N},
hydrostatic balance of the background state \eqnref{eqn:hydrostatic}, and the
continuity equation~\eqnref{eqn:compressibility}, we can rewrite the momentum and
energy equations in terms of the two velocity components, $u$ and $w$, and the reduced
Lagrangian pressure fluctuation, $\dP$,

\begin{eqnarray}
	\label{eqn:energy}
	i \omega \, \dP &=& c^2 \bvec{\nabla} \cdot \bvec{u} = c^2 \left(ik_xu + \der[w]{z}\right) \; ,
\\
	\label{eqn:u(dP)}
	u &=& \frac{\sigma^2\omega}{\omega^4-\sigma^4}\left[\der[]{z} + \frac{\omega^2 k_x}{\sigma^2} - \frac{1}{H}\right] \dP \; ,
\\
	\label{eqn:w(dP)}
	w &=& -\frac{i\omega^3}{\omega^4-\sigma^4} \left[\der[]{z} + \frac{\sigma^2 k_x}{\omega^2} - \frac{1}{H}\right] \dP \; .
\end{eqnarray}

\noindent Equations~\eqnref{eqn:u(dP)} and \eqnref{eqn:w(dP)} have
been obtained from the two components of the momentum equation by using the definition
of the Lagrangian pressure fluctuation~\eqnref{eqn:Lagrangian_pressure}, the modified
energy equation~\eqnref{eqn:energy}, and the continuity equation~\eqnref{eqn:compressibility}
to eliminate the Eulerian pressure fluctuation $P_1$, the density fluctuation $\rho_1$,
and the vertical derivative of the vertical velocity $dw/dz$. The resulting equations
have then been cross multiplied and subtracted to generate an equation that depends only
on the horizontal velocity component $u$ and another that depends only on the vertical
velocity $w$.

For compactness in these expressions, we have defined a frequency, $\sigma$,

\begin{equation}
	\label{eqn:sigma}
	\sigma^2 \equiv gk_x - 2\Omega\omega \; ,
\end{equation}

\noindent that generates the dispersion relation $\omega^2 = \sigma^2$ for surface
gravity waves, or $f$ modes, in our rotating star. The existence of such a solution
can be deduced from Equations~\eqnref{eqn:u(dP)} and \eqnref{eqn:w(dP)} by seeking
non-trivial solutions for the velocity components under the condition that the motions
are incompressible ($\dP = 0$). Due to the orthogonality of gravity and the rotation
vector at the equator, these solutions differ from the traditional Poincar\'e
waves---see Section~\ref{subsec:f-mode} for details.

Equations~\eqnref{eqn:energy}, \eqnref{eqn:u(dP)}, and \eqnref{eqn:w(dP)} can be
combined to generate an ODE whose dependent variable is the reduced Lagrangian
pressure fluctuation, $\dP$,

\begin{equation}
	\label{eqn:governing_equation}
	\left\{\dern[]{z}{2} - \frac{1}{H} \der[]{z}
		+ \left[ \frac{\omega^2 - 4\Omega^2}{c^2} + \frac{H'}{H^2} 
			- k_x^2 \left(1 - \frac{N^2}{\omega^2}\right)
			+ \frac{2\Omega k_x}{\omega}\left(\frac{1}{H} - \frac{2N^2}{g}\right)
				\right]\right\} \dP = 0 \; . 
\end{equation}

\noindent This governing equation is valid for a general stratification and for
a completely compressible fluid.  All of the atmospheric profiles are general
functions of height, i.e., $H = H(z)$ and $N^2 = N^2(z)$. For compactness
of notation, when convenient, we use a prime to indicate derivatives with respect
to the vertical coordinate $z$. Hence, $H' = dH/dz$ is the vertical derivative
of the density scale height.


\section{Local Dispersion Relation for a General Stratification}
\label{sec:Local_Dispersion} 

The governing equation~\eqnref{eqn:governing_equation} can be rewritten in standard
form (i.e., as a Helmholtz equation) by making a change of variable, $\dP(z) = \rho_0^{-1/2} \Psi(z)$,
that has been explicitly chosen to ensure that the resulting ODE for $\Psi(z)$ lacks
a first derivative term,

\begin{eqnarray}
	&& \dern[\Psi]{z}{2} + k_z^2(z) \Psi(z) = 0 \; ,
 	\label{eqn:standard_form}
\\
	&& k_z^2(z) \equiv \frac{\omega^2 - \left(\omega_c^2 + 4\Omega^2\right)}{c^2}
			- k_x^2 \left(1 - \frac{N^2}{\omega^2}\right)
			+ \frac{2\Omega k_x}{\omega\Hc} \; .
	\label{eqn:local_dispersion}
\end{eqnarray}

\noindent In the expression for $k_z^2$, the acoustic-cutoff frequency, $\omega_c$,
has the standard definition

\begin{equation}
	\omega_c^2 \equiv \frac{1 - 2H'}{4H^2} \, c^2 \; ,
\end{equation}

\noindent and $\Hc$ is a scale height that depends on the stratification,

\begin{equation}
	\frac{1}{\Hc} \equiv \frac{1}{H} -\frac{2N^2}{g} \; .
\end{equation}

\noindent In a stellar convection zone where $N^2 \approx 0$, the scale height $\Hc$ 
is nearly equal to the density scale height, but even in a region of stable stratification,
$\Hc$ is positive and in low-mass stars has the same magnitude as the density scale
height. Figure~\ref{fig:freq_ratio} illustrates the buoyancy frequency, the density
scale height $H$, and the scale height $\Hc$ as predicted by an evolutionary model
of the Sun's internal structure, i.e., Model S from \cite{Christensen-Dalsgaard:1996}. 

The quantity $k_z(z)$ is a local radial wavenumber that varies with height in the
atmosphere and Equation~\eqnref{eqn:local_dispersion} is a local dispersion relation
that describes acoustic waves and gravito-inertial waves. This local dispersion relation
reduces to the well-known expression for acoustic-gravity waves when the rotation
rate vanishes. The inclusion of the Coriolis force provides two modifications. First,
there is an effective correction to the acoustic-cutoff frequency (i.e.,
$\omega_c^2 \to \omega_c^2 + 4 \Omega^2$) that is exceedingly weak in most stars.
Second, the last term in Equation~\eqnref{eqn:local_dispersion} is solely due to the
Coriolis force and it is responsible for generating inertial waves.

The local dispersion relation constitutes a fourth-order polynomial equation in $\omega$
and hence for any given value of the vertical wavenumber, $k_z$, there are four solutions.
Two of the solutions correspond to high-frequency acoustic waves and two to low-frequency
gravito-inertial waves. In the limit of an isothermal atmosphere all of the atmospheric
profiles become constants, and hence the vertical wavenumber $k_z$ is a constant.
Figure~\ref{fig:isothermal_propagation} provides two propagation diagrams which illustrate
the frequencies for which the solutions are radially oscillatory in an isothermal atmosphere.
The two panels correspond to different values of the ratio of the rotation rate and the
buoyancy frequency. The shaded regions indicate the domains of vertical propagation where
the wavenumber $k_z$ is real valued ($k_z^2 > 0$). The white, unshaded regions correspond
to evanescent waves, where the wavenumber $k_z$ is purely imaginary ($k_z^2 < 0$) and the
solutions are exponentially growing or decaying with height. The high-frequency acoustic
waves fall into the orange portions of the diagram, with disjoint prograde and retrograde
solution branches. For an atmosphere with convectively stable stratification ($N^2 > 0$),
such as an isothermal atmosphere, the low-frequency gravito-inertial waves also possess
prograde and retrograde solution branches. In Figure~\ref{fig:isothermal_propagation} the
prograde solutions are shaded blue and the retrograde waves pink. The solid purple curve
is the bounding frequency between propagating and evanescent waves and is obtained by
setting $k_z = 0$ in the dispersion relation, Equation~\eqnref{eqn:local_dispersion}.
The dashed black horizontal lines indicate the inertial frequency range, $\omega = \pm 2 \Omega$
and the dotted black lines indicate the positive and negative values of the buoyancy
frequency $\omega = \pm N$. The two green dot-dashed curves correspond to the prograde
and retrograde $f$ modes.

Separate dispersion relations can be developed for the acoustic waves and the gravito-inertial
waves by considering high and low frequency limits of the full dispersion relation. In
the high-frequency limit, we ignore those terms that have the frequency $\omega$ in the
denominator, producing the following dispersion relation 

\begin{equation}
	\omega^2 \approx \left( k_x^2 + k_z^2 \right) c^2 + \omega_c^2+4\Omega^2 \; .
\end{equation}

\noindent The primary rotational modification to the $p$ modes is a slight increase in
the acoustic cutoff frequency that affects the prograde and retrograde solutions in the
same manner. The low-frequency gravito-inertial waves are obtained by neglecting the terms
$(\omega^2 - 4 \Omega^2)/c^2$,

\begin{equation}
	k_z^2(z) \approx -k_c^2 - k_x^2\left(1-\frac{N^2}{\omega^2}\right)
		+ \frac{2\Omega k_x}{\omega \Hc} \; ,
	\label{eqn:low-freq_dispersion}
\end{equation}

\noindent where we have defined a cutoff wavenumber, $k_c$, that depends purely
on the density stratification

\begin{equation}
	k_c^2(z) \equiv \frac{\omega_c^2}{c^2} = \frac{1 - 2H'}{4H^2} \; .
\end{equation}

\subsection{Low-Frequency Gravito-Inertial Waves}
\label{subsec:gravito-inertial_waves}

The low-frequency version of the local dispersion relation~\eqnref{eqn:low-freq_dispersion}
can be solved for the temporal frequency $\omega$ as long as we keep in mind that
all of the atmospheric profiles and the radial wavenumber are functions of height
but the frequency is a global, constant property. Equation~\eqnref{eqn:low-freq_dispersion}
is quadratic in the frequency; hence, there are two solutions given by

\begin{equation}
	\omega = \frac{k_x \Omega}{K^2 \Hc} \left[ 1 \pm \left(1 + K^2 \Hc^2 \frac{N^2}{\Omega^2}\right)^{1/2}\right] \; ,
	\label{eqn:frequency_relation}
\end{equation}

\noindent where we have defined a wavenumber $K$,

\begin{equation}
	K^2(z) \equiv k_x^2 + k_z^2(z) + k_c^2(z) \; ,
\end{equation}

\noindent that is related to the total wavenumber of the wave, but includes a
modification for stratification (through $k_c^2$). The nature of the two solutions
depends explicitly on the relative importance of the restoring forces of buoyancy
and the Coriolis force.  We will examine the possible extremes in the next two
subsections.

\subsubsection{Limit of Slow Rotation}
\label{subsubsec:slow_rotation} 

First, consider the situation where the magnitude of the buoyancy frequency is
enormous compared to the rotation rate, $\Omega/|N| \ll 1$.  In a low-mass star,
such conditions occur both above and below the convection zone, within the outer
atmosphere and within the radiative interior, as is illustrated in Figure~\ref{fig:freq_ratio}$a$.
In these regions, the ratio of frequencies, $\Omega/|N|$, can be treated as a small
parameter and we may expand the low-frequency dispersion relation \eqnref{eqn:low-freq_dispersion}
for slow rotation rate (or strong buoyancy), finding

\begin{equation}
	\omega = \pm \frac{k_x}{K} N + \frac{k_x \Omega}{K^2 \Hc} + \cdots \; .
	\label{eqn:slow_rotation}
\end{equation}

\noindent We have kept only the first two terms in the expansion and neglected all
terms of order $(\Omega/N)^3$ or smaller. In a stably stratified atmosphere ($N^2>0$),
the two solutions correspond to internal gravity waves that propagate in either
the prograde or retrograde directions. In an unstable stratification, both solutions
are unstable convective modes that propagate slowly prograde due to rotation. Of course,
the effects of stratification arise from the buoyancy frequency $N$, but they also
manifest through the wavenumber $K$ that appears in the denominator of the first
term in the expansion. Each wave has a small correction arising from the Coriolis
force that has the familiar form $mC\Omega$, where $C = 1/\left(K^2 R\Hc\right)$
is a parameter depending on the stratification and wavenumbers. Since the sign of
the correction is the same for both the prograde and retrograde solutions, the prograde
mode propagates slightly faster than the retrograde gravity wave.

Figure~\ref{fig:isothermal_propagation}$a$ provides a propagation diagram for an
isothermal atmosphere that is representative of the slow rotation limit (here with
$\Omega/N = 0.2$). The prograde and retrograde branches are only weakly asymmetric,
with the Coriolis force introducing the largest anisotropy at low zonal wavenumber,
$k_x$. As the zonal wavenumber becomes large, the propagation band approach that
expected for internal gravity waves $|\omega| < N$. We note that the expansion that
appears in Equation~\eqnref{eqn:slow_rotation} is valid not only in the limit of
slow rotation, but also in the limit of short zonal wavelength, $\Omega / |N| \ll K \Hc \approx k_x \Hc$.
Thus, for sufficiently large values of the zonal wavenumber, the buoyancy frequency
term will dominate and the gravito-inertial waves behave like internal gravity waves.

\subsubsection{Limit of Rapid Rotation}
\label{subsubsec:rapid_rotation}

The converse limit, where the rotation rate is much larger in magnitude than the
buoyancy frequency may hold in a star's convection zone where efficient convection
drives the unstable stratification towards neutral stability. Of course, the stratification
remains slightly unstable, $N^2 < 0$, and the buoyancy frequency itself is purely
imaginary. If we consider the limit of rapid rotation (or equivalently, weak buoyancy),
$\Omega / |N| \gg K \Hc$, we find two gravito-inertial waves with very
different natures. One is a prograde-propagating, almost-pure, inertial oscillation,
i.e., a thermal Rossby wave, with a frequency given by

\begin{equation}
	\omega = \frac{2\Omega k_x}{K^2 \Hc} + \frac{k_x \Hc}{2} \frac{N^2}{\Omega} + \cdots \; ,
	\label{eqn:dispersion_LowFreq_Prograde}
\end{equation}

\noindent where we have neglected all terms of order $(N/\Omega)^4$ or smaller.
The other solution is a slowly propagating gravito-inertial wave of mixed character,

\begin{equation}
	\omega = -\frac{k_x \Hc}{2} \frac{N^2}{\Omega} + \cdots \; ,
\end{equation}

\noindent where we have kept only the first nonzero term in the expansion. For a
stable stratification with $N^2>0$, like an isothermal atmosphere, the slow
gravito-inertial wave is retrograde propagating. For completeness, this situation
is illustrated in Figure~\ref{fig:isothermal_propagation}$b$, but we recognize
that this rapidly rotating limit never occurs in the stably stratified regions of
a star. In an unstable stratification where $N^2<0$, the slow gravito-inertial
wave is prograde. Hence, in a stellar convection zone, both the fast and slow
gravito-inertial waves are prograde with positive frequencies.

The reader should note that the slow gravito-inertial wave approaches zero frequency
in the limit of neutral stability ($N^2 \to 0$). Thus, in such an isentropic atmosphere,
one of the solution branches corresponds to a prograde thermal Rossby wave and the
other to a geostrophic mode that is stationary in the rotating frame of reference.

\subsection{Wave Cavities and Turning Points}
\label{subsec:wave_cavities} 

The gravito-inertial waves described by Equations~\eqnref{eqn:standard_form} and
\eqnref{eqn:low-freq_dispersion} are radially propagating wherever the local wavenumber
is purely real, $k_z^2>0$, and evanescent where it is imaginary, $k_z^2 < 0$. A
wave cavity therefore exists wherever $k_z^2>0$ and the boundaries of that cavity
correspond to the turning points of the equation where $k_z^2(z) = 0$. As we shall
soon see, for low-mass stars this can result in two potential cavities, one in
the convection zone and one in the radiative interior.  These cavities are in fact
waveguides, with waves trapped in radius and freely propagating in the zonal direction.

To see where a wave cavity exists, we need to consider the sign of each term in
the local dispersion relation. Since $\Hc > 0$ and $-k_c^2 < 0$ (the density scale
height $H$ decreases with radius in a star, so $H^\prime < 0$), only two of the terms
in the local dispersion relation are potentially positive. We rewrite the low-frequency
dispersion relation~\eqnref{eqn:low-freq_dispersion}, moving these two terms to the
front of the right-hand side, in order to emphasize which terms produce wave propagation,

\begin{equation}
	k_z^2 = \left[ \frac{2\Omega k_x}{\omega \Hc} + k_x^2 \frac{N^2}{\omega^2} \right]
		- \left(k_x^2 + k_c^2\right) \; .
\end{equation}

\noindent  For a wave cavity to exist, the term in the square brackets must exceed
$k_x^2 + k_c^2$. Clearly, this can happen for low, real-valued frequencies. The first
term, the Coriolis term, is large for frequencies less than the rotation rate and
is responsible for inertial wave cavities. In order for this term to be positive the
waves {\bf must be prograde} ($\omega/k_x > 0$). The second term, the buoyancy term,
is large for frequencies less than the buoyancy frequency and results in a cavity for
internal gravity waves. Both prograde and retrograde gravity waves are possible, but
the term leading to radial propagation is positive only when the atmosphere is stably
stratified $N^2 > 0$. If both terms are large, a cavity of mixed gravito-inertial waves
results. Finally, we note that even in an unstably stratified atmosphere, $N^2<0$, the
Coriolis term can dominate the buoyancy term for sufficiently long wavelengths and for
rapid enough rotation. These overstable convective modes have been stabilized by rotation
and have been suggested as a possible mechanism for the excitation of the pulsation
of $\beta$ Cephei stars \citep{Osaki:1974, Lee:1986, Lee:1987} through coupling to
gravity modes with similar frequency that reside in the overlying stably stratified
outer envelope.

Since the buoyancy frequency dominates the rotation rate in a star's radiative interior,
the resulting wave cavity is the well-known $g$-mode cavity with small corrections for
the Coriolis force. In the deep interior, the cutoff wavenumber is much smaller than any
reasonable zonal wavenumber, $k_c \ll k_x$. Hence, a bounding frequency
for radial propagation can be found by setting $k_z^2=0$ in the dispersion relation that
is valid for slow rotation, Equation~\eqnref{eqn:slow_rotation}. Under these conditions, 
$K^2 = k_x^2 + k_z^2 + k_c^2 \approx k_x^2$ and propagation occurs for frequencies below
the resulting bound,
 
\begin{equation}
	\left|\omega\right| < N \pm \frac{\Omega}{k_x \Hc} + \cdots \; .
\end{equation}	

\noindent The positive sign refers to the prograde $g$ mode and the negative sign to the
retrograde mode.

In a star's convection zone, if we can ignore the buoyancy frequency ($N^2=0$),
we expect a prograde thermal Rossby wave. We can easily demonstrate that a cavity
exists for sufficiently rapid rotation. For an isentropic stratification, the
scale height $\Hc$ reduces to the density scale height, $\Hc=H$. Further,
the density scale height is an increasing function of depth within a star (because
the temperature increases inwards). Therefore, the positive Coriolis term in the
dispersion relation, $2\Omega k_x/\omega H$, becomes small deep within the star
and at some depth below the photosphere there exists a lower turning point where,

\begin{equation}
	\frac{2\Omega k_x}{\omega H} \approx k_x^2 \;  ,
	\label{eqn:lower_turn}
\end{equation}

\noindent and at this depth, downward propagating thermal Rossby waves are refracted
back upwards towards the photosphere. Similarly, near the photosphere where $H$
becomes small and $k_c^2 = (1-2H')/4H^2$ becomes large, there is an upper
turning point where

\begin{equation}
	\frac{2\Omega k_x}{\omega H} \approx k_c^2 \; .
	\label{eqn:upper_turn}
\end{equation}
 
\noindent At this height an upward traveling inertial wave is reflected downwards.
The inertial waves are therefore trapped between these two turning points and the
region in between coincides with the wave cavity. The general condition for radial
propagation throughout a stellar convection zone can be expressed in terms of the
temporal frequency through the following inequality

\begin{equation}
	\omega < \frac{2 k_x \Omega}{\left(k_x^2+k_c^2\right) H} + \cdots \; .
\end{equation}


\section{Inertial Waves in a Neutrally Stable Stratification}
\label{sec:Neutrally_Stable}

Stellar convection is exceedingly efficient in transporting heat and the resulting
outward heat flux drives the stratification of the convection zone towards neutral
stability. Thus, a commonly adopted model for a stellar convection zone is an isentropic
atmosphere for which $N^2 = 0$. The buoyancy force vanishes in such an atmosphere
and the waves become pure inertial oscillations, greatly simplifying the behavior
of the wave field. Thus, we will examine inertial-wave propagation in an isentropic
stratification with some detail.

A neutrally stable atmosphere is a special case of a polytropic atmosphere. For our
purposes, it is sufficient to define a polytropic atmosphere as one that possesses
a constant vertical temperature gradient, $T_0' = -Q$, and a corresponding temperature
profile that is a linear function of height,

\begin{equation}
	T_0 = -Q z \; .
\end{equation}

\noindent In this expression, we have chosen to place the origin at the height where
the linear temperature profile vanishes. The polytropic atmosphere only exists in the
half-space below the origin, $z<0$, where the temperature is positive. For an ideal
gas in a plane-parallel atmosphere with constant gravity, such a temperature gradient
results in power-law relations for the thermodynamic variables,

\begin{eqnarray}
	\rho_0(z) &=& A \, (-z)^\alpha \; ,
	\label{eqn:poly_density}
\\
	P_0(z) &=& \frac{gA}{\alpha+1} (-z)^{\alpha+1} \; ,
\\
	c^2(z) &=& \frac{\gamma g(-z)}{\alpha+1} \; ,
\end{eqnarray}

\noindent where $A$ is an arbitrary constant and $\alpha$ is a dimensionless parameter
called the polytropic index that is related to the temperature gradient, $Q$,

\begin{equation}
	\alpha = \frac{g}{R_{\rm gas} Q} - 1 \; .
\end{equation}

\noindent The density scale height $H$, the buoyancy frequency $N$, and cutoff
wavenumber $k_c$ are similarly power laws,

\begin{eqnarray}
	H &=& \frac{(-z)}{\alpha} \; ,
	\label{eqn:poly_H}
\\
	N^2 &=& \frac{\alpha-\hat{\alpha}}{\gamma\hat{\alpha}} \frac{g}{(-z)} \; ,
	\label{eqn:poly_N2}
\\
	k_c^2 &=& \frac{\alpha(\alpha+2)}{4z^2} \; .
\end{eqnarray}

\noindent The quantity $\hat{\alpha}$ that appears in the expression for the
buoyancy frequency is the value of the polytropic index that corresponds to
neutral stability, $\hat{\alpha} \equiv \left(\gamma-1\right)^{-1}$. For a fully
ionized, monatomic gas with $\gamma = 5/3$ this corresponds to a polytropic index
of $\hat{\alpha} = 3/2$.

\subsection{Wave Cavity for a Neutrally Stable Polytrope}
\label{subsec:polytropic_cavity}

In the limit of neutral stability, the local dispersion relationship that describes
low-frequency inertial waves, Equation~\eqnref{eqn:low-freq_dispersion}, simplifies
significantly,

\begin{equation}
	k_z^2(z) = \frac{2\Omega k_x}{\omega H} - k_x^2 - k_c^2 =
		-\frac{2 \kappa k_x}{z} - k_x^2 - \frac{\alpha(\alpha+2)}{4 z^2} \; .
	\label{eqn:local_dispersion_polytrope}
\end{equation}

\noindent In the final expression $\kappa$ is a constant defined by

\begin{equation}
	\kappa \equiv \frac{\alpha \Omega}{\omega} \; .
	\label{eqn:definition_kappa}
\end{equation}

\noindent The turning points that demark the boundaries of the inertial wave cavity can
be found by setting $k_z^2=0$ in the local dispersion relation and solving the resulting
quadratic equation for the two roots in $z$,

\begin{equation}
	z_{turn} = -k_x^{-1} \left[ \kappa \pm \sqrt{\kappa^2 - \frac{\alpha(\alpha+2)}{4}}\right] \; .
	\label{eqn:zturn}
\end{equation}

\noindent The negative sign generates the upper turning point and the plus sign
the lower turning point. From this expression for the turning points, we can
derive a necessary condition for the existence of a wave cavity. Two real roots
must exist and this requires that the argument of the square root is positive,

\begin{equation}
	\omega < \sqrt{\frac{\alpha}{\alpha+2}} \; 2\Omega \; . 
\end{equation}

\noindent Therefore, for an inertial wave cavity to exist, the frequency must lie
below a cutoff that depends on both the rotation rate $\Omega$ and on the stratification
(through the polytropic index $\alpha$). For a neutrally stable polytrope with $\alpha = 1.5$
the frequency must satisfy $\omega \lesssim 1.3 \, \Omega$.

A convenient form of the local dispersion relation can be obtained by defining a
dimensionless depth, $\zeta = -2 k_x z$, that scales the vertical coordinate with
the zonal wavenumber. The factor of 2 is included purely for convenience
in subsequent equations. Using this spatial variable, the local dispersion relation
can be written in the following form,

\begin{equation}
	\frac{k_z^2(z)}{k_x^2} = \frac{4 \alpha}{\zeta} \frac{\Omega}{\omega} - 1 - \frac{\alpha(\alpha+2)}{\zeta^2} \; .
\end{equation}

\noindent As one can see, the zonal wavenumber drops out of the right-hand side. This
allows us to plot the propagation band as a function of a dimensionless frequency,
$\omega/\Omega$, and dimensionless depth, $\zeta$, for any zonal wavenumber, capturing
all of the behavior in a single diagram. As the zonal wavenumber changes, the shape of
the cavity does not change, but its vertical extent scales with the wavelength.
Figure~\ref{fig:polytropic_propagation} shows such a propagation diagram for an
$\alpha=1.5$ neutrally stable polytrope. All of the vertically propagating waves are
prograde thermal Rossby waves (blue-shaded region in Figure~\ref{fig:polytropic_propagation})
since the retrograde branch is degenerate at zero frequency for a neutrally stable
atmosphere.

In Figure~\ref{fig:polytropic_propagation}, the dashed red and orange curves illustrate
the effects of refraction and reflection in determining the turning points. Deep in
the atmosphere, the balance is that of Equation~\eqnref{eqn:lower_turn}, where the
Coriolis term balances $k_x^2$ in the local dispersion relation. The red curve is
given by solving this balance for the frequency,

\begin{equation}
	\omega \approx \frac{2\Omega}{k_x H} = -\frac{2\alpha\Omega}{k_x z} \;  .
	\label{eqn:bounding_frequency}
\end{equation}

\noindent From this expression it is obvious that waves of different frequency have
different turning points. Solving for the lower turning point, $z_{\rm lower}$, we obtain

\begin{equation}
	z_{\rm lower} \approx -k_x^{-1} \frac{2\alpha\Omega}{\omega} \; .
	\label{eqn:z_lower}
\end{equation}

\noindent In particular, low-frequency waves have a much deeper lower turning point,
and in the limit $\omega \to 0$, the lower turning point becomes infinitely deep.

Similarly, the orange curve in Figure~\ref{fig:polytropic_propagation} corresponds to
the frequencies obtained by balancing the Coriolis term with $k_c^2$---as in
Equation~\eqnref{eqn:upper_turn}---which is the balance that holds very close to the
outer surface. The resulting frequency and the concomitant upper turning point,
$z_{\rm upper}$ are given by

\begin{eqnarray}
	\omega \approx \frac{2\Omega k_x}{k_c^2 H} = -\frac{8\Omega k_x z}{\alpha+1} \; ,
\\
	z_{\rm upper} \approx -k_x^{-1} \frac{(\alpha+1) \omega}{8\Omega} \; .
\end{eqnarray}

\noindent Lower frequencies have cavities that reach closer to the surface. However,
since inertial waves are restricted to frequencies less than $2\Omega$, the upper
turning point can never extend very deeply, $z_{\rm upper} > -(\alpha+1)/4k_x$.

Since the depth of the lower turning point scales inversely with the zonal wavenumber---see
Equation~\eqnref{eqn:z_lower}, waves with a short zonal wavelength are trapped
just below the surface.  Hence, these waves do not sense the spherical geometry
of the star, nor the radial variation of its gravitational acceleration. We can
see that our a priori neglect of these two effects are justified a posteriori in
the limit of short zonal wavelengths.

\subsection{Analytic Solution for a Neutrally Stable Polytrope}
\label{subsec:analytic_solution}

When the atmosphere is polytropic, the standard form of the wave equation reduces
to a well-studied ODE, namely Whittaker's Equation. To see this, in
Equations~\eqnref{eqn:standard_form} and \eqnref{eqn:local_dispersion_polytrope}
make a change of variable to the nondimensional depth that we introduced earlier,
$\zeta \equiv -2 k_x z$,

\begin{equation} \label{eqn:Whittaker}
	\dern[\Psi]{\zeta}{2} + 
		\left[ \frac{\kappa}{\zeta} - \frac{1}{4} + \frac{1/4-\mu^2}{\zeta^2} \right] \Psi(z) = 0 \; ,
\end{equation}

\noindent with the definition

\begin{equation}
	\mu \equiv \frac{\alpha+1}{2} \;  .
	\label{eqn:definition_mu}
\end{equation}

\noindent Note, both $\kappa$ and $\mu$ are constants. The constant $\kappa$ will
serve as the eigenvalue of the ODE and it depends on the wave frequency---see
Equation~\eqnref{eqn:definition_kappa}. We note that the definition
of $\kappa$ that appears in Equation~\eqnref{eqn:definition_kappa} is only valid
is the low-frequency regime of the local dispersion
relation~\eqnref{eqn:local_dispersion_polytrope}. The constant $\mu$ is a parameter
that depends purely on the stratification. Equation~\eqnref{eqn:standard_form} is
Whittaker's Equation \citep{Abramowitz:1964} which has two solutions called Whittaker
functions, ${\cal M}_{\kappa\mu}\left(\zeta\right)$ and ${\cal W}_{\kappa\mu}\left(\zeta\right)$.
These Whittaker functions can be expressed in terms of Kummer's confluent
hypergeometric functions of the first and second kind, $M$ and $U$ (confusingly,
sometimes referred to as Kummer's function and Tricomi's function, respectively),

\begin{eqnarray}
	{\cal M}_{\kappa\mu}\left(\zeta\right) = e^{-\zeta/2} \; \zeta^{\mu+1/2} \; M\left(-\eta, 1+2\mu, \zeta\right) \; ,
\\
	{\cal W}_{\kappa\mu}\left(\zeta\right) = e^{-\zeta/2} \; \zeta^{\mu+1/2} \; U\left(-\eta, 1+2\mu, \zeta\right) \; ,
\end{eqnarray}

\noindent where $\eta \equiv \kappa - \left(\mu + 1/2\right)$. The general solution
for the Lagrangian pressure fluctuation is therefore a linear combination of these
two solutions,

\begin{equation}
	\delta\varpi(z) = \rho_0^{-1/2} \, \Psi(z) = z \, e^{k_x z}
		\left[ C_a M\left(-\eta, \alpha+2, -2k_x z\right) +
			C_b U\left(-\eta, \alpha+2, -2k_x z\right)\right] \; ,
\end{equation}

\noindent with arbitrary constants $C_a$ and $C_b$ whose ratio is determined by
the boundary conditions.

\subsection{Boundary Conditions and Eigenmodes}
\label{subsec:eigenmodes}

When physically appropriate boundary conditions are applied in height, Whittaker's
Equation generates a discrete spectrum of eigenmodes with $\kappa$ serving as the
eigenvalue. Since $\kappa$ is a function of the frequency, each eigenmode will possess
a specific eigenfrequency,

\begin{equation}
	\omega_n = \frac{\alpha\Omega}{\kappa_n} \; ,
	\label{eqn:eigenfrequencies}
\end{equation}

\noindent where $n$ is the radial order of the mode and $\kappa_n=\kappa_n(k_x)$
is the $n$th eigenvalue which is potentially a function of the zonal wavenumber.

\subsubsection{Eigenmodes for a Semi-Infinite Polytrope}
\label{subsubsec:semi-infinite_layer}

Whittaker's equation has two singular points, one at the origin $\zeta = 0$ (i.e.,
at the upper surface of the polytrope, $z=0$) and the other at infinity $\zeta \to \infty$
(or $z \to -\infty$). In this subsection, we will model the convection zone of a
star as a semi-infinite polytropic layer that extends between these two singular
points. Obviously, a star's convection zone has a finite depth. But, as we have
already discussed, the wave cavity is confined to a region near the upper surface
and the depth to which this cavity extends is proportional to the zonal wavelength
of the wave. Thus, if the zonal wavelength is sufficiently short, the lower turning
point of the wave may be many e-folding lengths from the physical boundary at the
bottom of the convection zone. Hence, such waves do not sense the bottom and we can
treat the domain as semi-infinite. In the subsequent subsection, \S\ref{subsubsec:finite_layer},
we will consider what happens when the eigenfunctions begin to reach deeply enough
to be influenced by the lower boundary.

If we apply boundary conditions of regularity at the origin and at infinity (at
the two singular points), the solution simplifies dramatically. The $U$ confluent
hypergeometric function is badly divergent at the origin, so we reject it. The
$M$ confluent hypergeometric function is divergent at infinity unless the eigenvalue
$\kappa$ takes on specific discrete values. Explicitly, the parameter $\eta$ must
be a non-negative integer,

\begin{equation}
	\eta_n = \kappa_n - \left(\mu + 1/2\right) = n \; ,  \qquad\qquad   n\in 0,1,2,3, \ldots \; .
	\label{eqn:dispersion_eta}
\end{equation}

\noindent For these integer values of $\eta$, the Kummer functions reduce to associated
Laguerre polynomials and, to within an arbitrary constant amplitude, $C_n$, the eigenfunctions
become,

\begin{equation}
	\delta\varpi_n(z) = C_n \, z \, e^{k_x z} \, L_n^{(\alpha+1)}\left(-2k_x z\right) \; ,
\end{equation}

\noindent where $L_n^{(\alpha+1)}$ is the $n$th-order associated Laguerre polynomial.

For these boundary conditions the eigenfrequencies have the unexpected property
that they are insensitive to the zonal wavenumber. Inserting
Equation~\eqnref{eqn:dispersion_eta} into Equation~\eqnref{eqn:eigenfrequencies},
we obtain a global dispersion relation that expresses the mode frequency as a
function of only the radial order $n$, the rotation rate $\Omega$, and the polytropic
index $\alpha$,

\begin{equation}
	\omega_n = \frac{2\Omega}{1 + 2\left(n+1\right)/\alpha} \; .
	\label{eqn:omega_n}
\end{equation}

\noindent In Section~\ref{subsec:self-similar} we discuss in more detail why the
eigenfrequencies for this set of boundary conditions are independent of the zonal
wavenumber. But, in short, this property is the result of the self-similarity of
a polytropic atmosphere and the boundary conditions; neither possesses an imposed
spatial scale. In the next subsection, we will consider a finite spatial domain
and discover that the self-similarity of the boundary conditions is broken and the
eigenfrequencies will subsequently depend on the zonal wavenumber.
 
The eigenfrequencies are indicated in Figure~\ref{fig:polytropic_propagation} by
plotting a horizontal dotted blue line at the frequency corresponding to each eigenmode.
Each line extends between the two turning points, thus illustrating the spatial extent
of the wave cavity for each eigenmode. As is typical for inertial oscillations, the
lowest radial orders $n$ possess the highest frequencies and there is an accumulation
point at zero frequency as the mode order becomes large ($n\to\infty$).

Figure~\ref{fig:polytropic_eigenfunctions} shows radial eigenfunctions for modes
corresponding to the lowest four radial orders. The left panel shows the eigenfunction
for the Lagrangian pressure fluctuation, scaled by $\omega/g$ such that the eigenfunctions
have physical units of velocity. The right-hand panel of Figure~\ref{fig:polytropic_eigenfunctions}
presents the corresponding eigenfunctions for the zonal velocity, Equation~\eqnref{eqn:u(dP)}.
We do not illustrate the vertical velocity separately, because the vertical velocity
is nearly proportional to the reduced Lagrangian pressure fluctuation when in the
low-frequency limit, see Section~\ref{subsec:anelasticity}. In fact, in this low-frequency
limit, the left-hand panel plots the imaginary component of the vertical velocity as well
as the Lagrangian pressure fluctuation, as can be verified by taking the low-frequency
limit of Equation~\eqnref{eqn:w(dP)}. Each of the eigenfunctions has been
normalized such that the zonal velocity has a maximum amplitude of unity that is
achieved at the upper surface. In general, the zonal velocity is the larger of the
two velocity components, but this is especially true near the upper surface. The
vertical velocity vanishes at the upper boundary for all modes, while the zonal
velocity reaches maximum amplitude at the surface. This is a result of the upper
surface being a regular singular point of Whittaker's equation. 

The wave cavity is different for each mode and extends from the lower turning
point---which is indicated in Figure~\ref{fig:polytropic_eigenfunctions} by the
solid circular dot---to the upper turning point which lies just below the upper
boundary (not illustrated for clarity). For the four radial modes that are shown,
the upper turning points all lie very close to the upper surface and very near
each other, with $k_x z_{\rm upper}$ ranging between $-0.43$ and $-0.14$.

\subsubsection{Eigenmodes for a Polytropic Layer of Finite Depth}
\label{subsubsec:finite_layer} 

As can be seen in Figure~\ref{fig:polytropic_eigenfunctions} as the zonal wavelength
increases or the frequency decreases, the eigenfunctions reach deeper and deeper
into the star. Eventually, the eigenfunction becomes sensitive to the bottom of
the convection zone. To test such effects we consider a finite layer that spans
the range of heights $z\in[-D,\,0\,]$, where $D=200$ Mm is approximately the depth
of the Sun's convection zone. Hence, boundary conditions are applied at the singular
point corresponding to the origin and at the regular point $z=-D$. As before, we
require regularity at the upper surface. For simplicity, we require
that the Lagrangian pressure fluctuation vanishes at the bottom of the convection
zone, $\dP(-D)=0$. This boundary condition is consistent with one of impenetrability
$w=0$ at low frequencies. As before, the boundary condition of
regularity at $z=0$, forces us to reject the $U$ confluent 
hypergeometric function. The boundary condition at the bottom of the convection zone
then generates a transcendental global dispersion relation involving the $M$ Kummer
function,

\begin{equation}
	M\left(-\eta, \, \alpha+2, \, 2 k_x D\right) = 0 \; .
\end{equation}

\noindent We have solved this equation numerically for the discrete set of eigenvalues
($\eta = \eta_n$), and the results are shown in Figure~\ref{fig:finitedepth_eigfreqs}.
The solid black, red, blue, and green curves illustrate the mode frequencies for the
radial fundamental $n=0$ (black) and the first three overtones $n=1$, 2, and 3 (red,
blue and green respectively). The light-blue curves represent all of the higher order
overtones, which have not been labeled for the sake of clarity. As expected there is
an accumulation point at zero frequency as $n\to \infty$, and this causes all of the
pale blue curves to appear to merge into a solid band of color.

Two specific eigenmodes are marked along each of the first four dispersion curves.
The diamond symbol corresponds to a mode with low wavenumber $m = k_x R_\sun = 10$
and the circular symbol to a higher wavenumber $m = 40$. The eigenfunctions
for these modes are shown in Figure~\ref{fig:finitedepth_eigfuncs}. The upper panels
show the Lagrangian pressure fluctuation and the zonal velocity for the smaller wavenumber
($m = 10$), while the lower panels illustrate the same quantities for the larger wavenumber
($m=40$).

At low values of the zonal wavenumber, the frequency along a single dispersion curve
increases linearly with wavenumber. This is the wavenumber regime in which the lower
turning point of the wave equation lies below the lower boundary, $z_{\rm lower} < -D$.
Thus, the eigenfunctions extend throughout the entirety of the finite domain, and the
lower boundary condition has  significant influence on the eigenmodes. The modes marked
with the diamond symbols in Figure~\ref{fig:finitedepth_eigfreqs} are all within this
regime, and one can see that their eigenfunctions remain propagating all the way to
the bottom of the spatial domain.

As the wavenumber becomes larger, the lower turning point eventually crosses the
lower boundary and passes into the solution domain. We can approximate the frequency
for which the lower turning point and the lower boundary are coincident
($z_{\rm lower} = -D$) by using Equation~\eqnref{eqn:z_lower},

\begin{equation}
	\frac{\omega}{\Omega} \approx \frac{2\alpha}{k_x D} \; .
\end{equation}

\noindent This transition frequency is shown in Figure~\ref{fig:finitedepth_eigfreqs}
using the dashed purple curve. Further increase to the wavenumber results in the
withdrawal of the wave cavity from the lower boundary and the presence of that
lower boundary becomes less and less important to the eigenmode. Hence, for sufficiently
large wavenumber (those that lie to the right of the dashed purple curve), the
dispersion curves in Figure~\ref{fig:finitedepth_eigfreqs} asymptote to the constant
frequencies associated with eigenmodes of the semi-infinite domain. These asymptotic
values, Equation~\eqnref{eqn:omega_n}, are shown with the dotted lines. For the sake
of clarity, dotted lines are shown only for the first seven modes. The eigenmodes
that are indicated by the circular symbols all lie within this high-wavenumber
regime and possess lower turning points that lie between the two boundaries. As
such, their wave cavities do not extend all the way to the bottom of the convection
zone. To illustrate this,  Figure~\ref{fig:finitedepth_eigfuncs}$d$ indicates the
lower turning point for each mode with an appropriately colored circular dot.

\subsubsection{Eigenmodes for a Submerged Polytropic Layer}
\label{subsubsec:submerged_layer} 

In this subsection we consider a buried layer of finite radial extent that is
fully submerged beneath the polytrope's singular upper surface. We place impenetrable
boundaries at two depths, $z=-D$ and $z = -d$, with $D > d$. Thus as in the previous
subsection, the lower boundary is at $-D$, but now instead of being located at the
origin, the upper boundary is at $-d$. Since, neither of the ODEs singular points
lie within the spatial domain, $z \in [-D,-d]$, we must retain both of the confluent
hypergeometric functions and we must find the linear combination of $M$ and $U$ that
causes the Lagrangian pressure fluctuation (or, equivalently at low frequencies,
the vertical velocity) to vanish at both boundaries. This is readily accomplished
by using a numerical root finder to obtain the roots of the following matrix determinant,

\begin{equation}
  	\left| {\begin{array}{cc}
   		M\left(-\eta, \alpha+2, 2k_x D\right) & U\left(-\eta, \alpha+2, 2k_x D\right) \\
   		M\left(-\eta, \alpha+2, 2k_x d\right) & U\left(-\eta, \alpha+2, 2k_x d\right) \\
  		\end{array} }
	\right| = 0 \; .
\end{equation}

Figure~\ref{fig:finitedomain_eigfreqs} provides the resulting dispersion diagram
for a model with $d = 30$ Mm and $D = 200$ Mm. The depression of the upper boundary
condition below the singularity of the polytrope modifies the asymptotic behavior
for large wavenumbers. Instead of approaching a constant value, the frequencies
monotonically decrease after achieving a local maximum which lies near the wavenumber
marking the transition from low- to high-wavenumber regimes in
Figure~\ref{fig:finitedepth_eigfreqs}. This behavior arises because the size of
the wave cavity is reduced for those waves that have an upper turning point that
lies above the upper boundary $z_{\rm upper} > -d$. In order to fit the same number
of vertical wavelengths within the reduced domain, the frequency must be decreased.
Further, the truncation of the wave cavity is more severe for waves with a large
zonal wavenumber because the cavity is fundamentally thinner. This same frequency
behavior was found previously by \cite{Glatzmaier:1981c}, but they solved the ODE
by Frobenius expansions and presented solutions only for the radial fundamental $n=0$.


\section{Discussion}
\label{sec:Discussion}

Under the assumptions of short longitudinal wavelengths and 2D motions perpendicular
to the rotation axis, we have derived an ODE, Equation~\eqnref{eqn:governing_equation},
that describes the propagation of atmospheric waves through an atmosphere with
general stratification. There are two families of solutions: a high-frequency branch
of acoustic waves and a low-frequency branch of gravito-inertial waves. From this
governing ODE, we have derived a local dispersion relation and have demonstrated
that in the absence of buoyancy the thermal Rossby waves are trapped radially in
a waveguide that is confined to the upper reaches of the convection zone. The upper
boundary of this waveguide is caused by reflection of an upward propagating wave
by the density stratification---specifically, when the vertical wavelength becomes
comparable to the density scale height. This reflection is substantiated in the
dispersion relation by a cutoff wavenumber that is related to the acoustic cutoff
frequency. The lower boundary of the waveguide is caused by the upward refraction
of obliquely propagating wave fronts as the wave propagates downward into a region
with increasing density scale height. Due to the refractive nature of this turning
point, waves with shorter longitudinal wavelengths (large $k_x$) have a shallower
wave cavity and are trapped closer to the upper boundary.

\subsection{Location of the Waveguide}
\label{subsec:location_waveguide}

The exact location of the cavity for thermal Rossby waves is rather sensitive to
the superadiabatic gradient. We have shown that for a neutral stratification,
$N^2 = 0$, the cavity is in the upper portion of the convection zone. This result
is robust as long as the magnitude of the buoyancy frequency is small compared
to the rotation rate. Unfortunately, this condition is unlikely to hold true for
the solar-like stars. Figure~\ref{fig:freq_ratio} presents the buoyancy frequency
in a standard model of the Sun's internal structure, i.e., Model S of
\cite{Christensen-Dalsgaard:1996}. The left-hand panel shows the square of the
buoyancy frequency throughout the Sun's interior (blue curve). The convection zone
corresponds to the region where the square of the buoyancy frequency is negative.
For reference, the Sun's Carrington rotation rate $\Omega_\sun = 2.87 \times 10^{-6}$ s$^{-1}$
is indicated by the two red horizontal lines (located at $\omega^2 = \pm \Omega_\sun^2$).
The figure suggests that the upper and lower halves of the convection zone are in
converse regimes. The upper convection zone is in the slow rotation limit where
$\left|N\right| \gg \Omega$ and the deeper portion of the convection zone is in the
limit of rapid rotation, $\left|N\right| \ll \Omega$. If this suggestion is true,
the upper portion of the convection zone is inherently unstable for waves of all
wavelengths, but the lower portion of the convection zone might support stable thermal
Rossby waves, or equivalently overstable convective modes, if their wavelengths are
sufficiently long. How long?  An estimate can be derived from the local dispersion
relation---Equation~\eqnref{eqn:frequency_relation}, by requiring that the temporal
frequency is purely real. When the square of the buoyancy frequency is negative, this
provides a restriction on the wavenumber,

\begin{equation}
	k_x^2 + k_z^2 < \frac{\Omega^2}{\Hc^2\left|N^2\right|} - k_c^2 \; .
\end{equation}

\noindent Deep in the convection zone we can ignore $k_c^2$ and $\Hc \approx H$,
resulting in the following expression for the wavelength, $\lambda$,

\begin{equation}
	\frac{\lambda}{2\pi} = \left(k_x^2 + k_z^2\right)^{-1/2} =  H \frac{\left|N\right|}{\Omega} \; .
\end{equation}

\noindent From Figure~\ref{fig:freq_ratio}, we estimate that at the base of the
convection zone the ratio of the buoyancy frequency to the rotation rate has a
typical value of $\left|N\right| / \Omega \sim 0.1$. Further, at the base of the
convection zone the density scale height is roughly 100 Mm. Therefore, for
local stability, the gravito-inertial waves must have wavelengths longer than
60 Mm ($\sim 2\pi \times 0.1 \times 100~{\rm Mm}$). For waves with
similar longitudinal and vertical wavelengths, such waves would have azimuthal
orders $m \lesssim 30$.

The existence of a deep cavity for overstable convective modes in the Sun remains
rather speculative for a variety of reasons. First, the wavelength bound that we
just estimated is sufficiently large that our short wavelength approximation is
beginning to fray. Second, the stability condition is purely a local one. Calculating
global stability would require solving the eigenvalue problem for a solar-like
stratification. Third, while the sound speed and density profile of Model S (and
other solar models) have been very well tested by helioseismology, the superadiabaticity
in the convection zone remains poorly constrained. Helioseismology can only place
a rather large upper limit, and the buoyancy frequency could be much smaller in
magnitude and remain consistent with observations. In most solar models, the
superadiabaticity is completely determined by mixing-length theory. The entropy
gradient is fixed by the requirement that the convective heat flux carries a solar
luminosity through the convection zone. Thus, the modeled buoyancy frequency in
the convection zone is only as reliable as mixing-length theory is in detailed
modeling of the convective heat transport.

\subsection{The Anelastic Limit}
\label{subsec:anelasticity}

Even though internal gravity waves are usually much lower in frequency than acoustic
waves, compressibility is still important for their propagation.  Hence, anelastic
treatments of the continuity equation, where the mass flux is assumed to be divergenceless,
$\bvec{\nabla} \cdot \left(\rho_0 \bvec{u}\right) \approx 0$, often cause internal
gravity waves to fail to conserve energy \citep{Brown:2012}. This is the primary
reason that here we have adopted a continuity equation appropriate for a completely
compressible fluid. Despite these well-known issues for internal gravity waves, one
still hopes that anelasticity should hold in the limit of slowly evolving motions,
i.e., for low frequencies. Satisfyingly, this proves to be true. If we assume that
the waves of interest have very low frequency, $\omega^2 \ll gk_x$, we may consider
limits of Equations~\eqnref{eqn:u(dP)} and \eqnref{eqn:w(dP)} under the condition
that $\omega^2 \ll \sigma^2$. To leading order we obtain,

\begin{eqnarray}
	u &\approx& -\frac{\omega}{gk_x - 2\Omega\omega} \left(\der[]{z} - \frac{1}{H}\right) \dP \; ,
	\label{eqn:low-freq_u}
\\
	w &\approx& \frac{i\omega k_x}{gk_x - 2\Omega\omega} \, \dP \; .
	\label{eqn:low-freq_w}
\end{eqnarray}

\noindent If one works their way back through the derivation of Equations~\eqnref{eqn:u(dP)}
and \eqnref{eqn:w(dP)}, one can determine that this low-frequency approximation is
equivalent to neglecting the inertial term that appears on the left-hand side of the
momentum equation~\eqnref{eqn:vector_momentum}. These equations demonstrate that very
low-frequency waves can be expressed using a stream function for the mass flux,

\begin{eqnarray}
	\rho_0 u &\approx& -\der[\psi]{z} \; ,
	\label{eqn:stream_u}
\\
	\rho_0 w &\approx& \der[\psi]{x} \; ,
	\label{eqn:stream_w}
\\
	\psi &\equiv& \frac{\omega}{gk_x - 2\Omega\omega}  \, \rho_0 \dP \; ,
\end{eqnarray}

\noindent where the stream function is proportional to the reduced Lagrangian
pressure fluctuation. Therefore, the mass flux must be divergenceless
$\bvec{\nabla} \cdot \left(\rho_0 \bvec{u}\right) \approx 0$, and the anelastic
approximation is satisfied. Further, by using Equation~\eqnref{eqn:Lagrangian_pressure}
to express the Lagrangian pressure fluctuation in terms of the Eulerian fluctuation,
one can demonstrate the these very low-frequency anelastic flows are geostrophic
to leading order,

\begin{eqnarray}
	\rho_0 u &\approx& \frac{1}{2\Omega}\der[P_1]{z} \; ,
\\
	\rho_0 w &\approx& -\frac{1}{2\Omega}\der[P_1]{x} \; .
\end{eqnarray}

\noindent For very low, but nonzero frequencies, inertia and buoyancy appear as
higher-order perturbations to geostrophic balance, and its the dynamics of these
perturbations that lead to prograde propagation of the flow pattern.

\subsection{Geometry of the Eigenfunctions}
\label{subsec:eigenfunction_geometry}

The eigenfunctions that we illustrate in Figures~\ref{fig:polytropic_eigenfunctions}
and \ref{fig:finitedepth_eigfuncs} correspond to a longitudinal sequence of rolls
aligned with the rotation axis. In order to illustrate this geometry, we provide
Figure~\ref{fig:polytropic_cells}, which shows the 2D motion for four distinct
eigenfunctions. The upper panels correspond to the fundamental radial eigenmode,
$n=0$, and the bottom panels to the first radial overtone, $n=1$. The right and
left panels provide eigenfunctions for two different zonal wavenumbers, a small
wavenumber $m=k_x R_\sun = 40$ on the left and a larger wavenumber $m=80$ on the
right. In each panel, longitude runs from right to left and the outer surface of
the star's convection zone lies at the top with the center of the star located
downwards beyond the bottom of each panel. The reader's viewpoint is consistent
with looking down on a piece of the star's equatorial plane from above the north
pole.

The colored images in Figure~\ref{fig:polytropic_cells} illustrate the stream
function, $\psi$, for the eigenfunction's mass flux. The black curves mark the
isocontours of the stream function and, hence, indicate the flowlines. Red tones
and dashed isocontours correspond to negative values of the stream function resulting
in clockwise flow along the isocontours (anticyclonic motion). Conversely, the
blue tones and solid curves indicate positive values of the stream function and
counter-clockwise, cyclonic flow. The thick black lines indicate the zero contours,
or the boundaries between cells. The flow field is composed of parallel rolls whose
axes are all aligned with the star's rotation vector. A single zonal wavelength
consists of two counter-rotating rolls, one clockwise and the other counter-clockwise.

As expected, the fundamental mode consists of a single roll in depth while the
first radial overtone possesses two counter-rotating rolls stacked in depth. The
upper roll looks weak in these renditions. This is an artifact of the stream function
generating the mass flux instead of the velocity field itself.  Since, the mass
density vanishes at the upper surface, so too does the mass flux.

The stream function for the fundamental mode $n=0$ (Figures~\ref{fig:polytropic_cells}$a$,$b$)
should be compared to the results of the numerical simulation illustrated in
Figure~\ref{fig:thermal_Rossby}$b$. Here and in the anelastic simulations
shown in Figure~\ref{fig:thermal_Rossby}, the vertical elongation of the vortical
columns is a function of the density stratification. The most obvious difference
between the eigenfunctions derived here and the numerical simulation is the cellular
tilt. The numerical simulations demonstrate that the upper portion of each cell is
centered at a higher longitude than the lower portion. Such tilting
behavior, or in extreme cases spiralling behavior, is a common feature that appears
in diffusive solutions in spherical geometry \citep{Zhang:1992, Yano:1992, Jones:2000,
Jones:2009}. Neither of these effects are present in the calculations that we perform
here.

\subsection{Self-Similarity of a Polytropic Atmosphere}
\label{subsec:self-similar}

In Section~\ref{subsubsec:semi-infinite_layer} we find that the eigenfrequencies
for inertial oscillations within a semi-infinite isentropic atmosphere lack dependence
on the zonal wavenumber. This unexpected result is due to the self-similarity of
an infinitely deep polytropic atmosphere. All of the thermodynamic profiles (temperature,
pressure, density, density scale height, etc.) are power-law functions of the depth
and lack an imposed spatial scale. Hence, when trying to nondimensionalize the ODE
that describes inertial waves within a polytropic atmosphere, the only length scale
that is available is the zonal wavelength. There is only one way to form a nondimensional
depth $\zeta$ (the independent variable of the governing equation), and it must be
proportional to $-k_x z$. The resulting non-dimensional form of the governing equation
is thus self-similar and so too are its solutions. This is the reason that the
eigenfunctions of the semi-infinite polytrope have the same shape, independent of
the zonal wavenumber, but their spatial extent in all directions scales linearly with
the zonal wavelength. For example, consider the eigensolutions illustrated in
Figure~\ref{fig:polytropic_cells}; the shape of the rolls in the equatorial plane
is independent of the longitudinal wavelength (compare the right and left panels).
The only difference is their spatial scale. Waves with a short longitudinal wavelength
have a short vertical wavelength, and vice versa.

The only way to break this self-similarity is by the imposition of a spatial scale
through the boundary conditions. For a semi-infinite atmosphere, with boundary
conditions placed at the origin and infinity, there is no imposed depth. Hence,
the boundary conditions are also self-similar and the resulting eigenfrequencies
can only depend on the stratification (through the polytropic index) and on the
rotation rate.  However, when boundaries are placed at finite depths within the
polytropic atmosphere, the depth of the domain provides a length scale and the
eigenfunctions and eigenfrequencies can now depend on the wavenumber. For example,
if we place boundaries at the origin and at a depth of $z=-D$, as we did in
Section~\ref{subsubsec:finite_layer}, the eigenfrequencies can now depend on the
zonal wavenumber through the combination $k_x D$.

\subsection{The $f$ Mode}
\label{subsec:f-mode}

From Equations~\eqnref{eqn:u(dP)} and \eqnref{eqn:w(dP)} one can deduce that there
is a solution that is incompressible and has vanishing Lagrangian pressure fluctuation
everywhere $\dP = 0$. In order to avoid the trivial solution, the dispersion relation
must satisfy $\omega^4 = \sigma^4$. The negative root, $\omega^2 = -\sigma^2$ corresponds
to an unphysical solution with an unbounded energy density. The positive root, 
$\omega^2 = \sigma^2$ leads to the traditional $f$ mode, or surface gravity
wave, that has the same exponential behavior as the $f$ mode for a non-rotating star,
but a modified eigenfrequency,

\begin{eqnarray}
	w = -i u &=& \exp\left({k_x z}\right) \, e^{ik_x x} \, e^{-i \omega t}  \; ,
\\
	\omega &=& \pm\left(gk + \Omega^2\right)^{1/2} - \Omega \; .
\end{eqnarray}

\noindent This dispersion relation is rather different from that of Poincar\'e
waves because in our model the gravitational acceleration and the rotation vector
are orthogonal, while in the traditional derivation of Poincar\'e waves the two
vectors are parallel.

In the limit of slow rotation, we recover the traditional dispersion relation for
deep water waves but with a rotational correction,

\begin{equation}
	\omega = \pm\sqrt{gk} - \Omega + \cdots \; .
\end{equation}

\noindent However, for rapid rotation, there is a fast retrograde solution that
is primarily an inertial oscillation

\begin{equation}
	\omega = -2\Omega  - \frac{gk}{2\Omega} + \cdots \; ,
\end{equation}

\noindent and a slow prograde solution that is a mode of strongly mixed character,

\begin{equation}
	\omega = \frac{gk}{2\Omega} + \cdots \; .
\end{equation}

Figure~\ref{fig:isothermal_propagation} presents the frequency of both the prograde
and retrograde surface gravity waves using the dot-dashed green curves. When the
rotation is strong, the anisotropy between the prograde and retrograde solutions
can be rather extreme.  In the right panel (rapid rotation), the prograde $f$ mode's
frequency is nearly equal to the upper frequency bound for the gravito-inertial
waves, while the retrograde $f$ modes shadows the boundary of the retrograde acoustic
waves.

\subsection{Observability of Thermal Rossby Waves in the Sun's Convection Zone}
\label{subsec:observability}

Recent observations of inertial oscillations in the Sun have been of three types:
classical equatorially confined Rossby waves \citep{Loeptien:2018, Liang:2019,
Proxauf:2020, Hanasoge:2019, Alshehhi:2019, Hanson:2020, Hathaway:2021}, critical-latitude
inertial modes \citep{Gizon:2020,Gizon:2021}, and high-latitude inertial modes
\citep{Bogart:2015, Gizon:2021, Hathaway:2021}. Thermal Rossby waves have not yet
been observed. Why is this when thermal Rossby waves are such a prominent feature
in laboratory experiments and in numerical simulations of the solar convection zone?

We suggest two related possibilities: the thermal Rossby waves are all unstable
or the potentially stable long-wavelength waves have a cavity that is too deep
to easily measure the waves at the solar surface. Convection in a star like the
Sun, and in fact all stars, is highly supercritical. So, the thermal Rossby waves
that appear at convective onset are not those that we should expect to see in solar
observations. Instead, due to the extreme level of turbulence, we should expect
to see the highly nonlinear cousins of those thermal Rossby waves that appear under
more laminar conditions. Surprisingly, numerical simulations suggest that the spatial 
scale of the most unstable thermal Rossby wave persists in highly turbulent regimes
\citep{Hindman:2020a}. This spatial scale is often associated with what are termed
convective banana cells \citep[e.g.,][]{Wilson:1988, Miesch:2000, Hotta:2015, Nelson:2018}.
Despite the fact that banana cells are immediately obvious to the eye in movie
sequences from convection simulations, they do not form an obvious feature in spectra
when the fluid is sufficiently turbulent \citep[e.g.,][]{Hindman:2020a}. The power
from these nonlinear waves is broadly spread across azimuthal order $m$, harmonic
degree $\ell$, and temporal frequency $\omega$. Thus, the power does not form a clean
dispersion relation that would verify its wavelike nature. One can deduce that the
convective features propagate prograde relative to the differential rotation, but
extracting more information has proved problematic.

Further, we must admit that it is uncertain what spatial scale should typify
thermal Rossby waves in the Sun. The convective columns that appear in numerical
simulations possess zonal wavenumbers that depend on the convection's Rossby
number Ro \citep{Featherstone:2016}. Unfortunately, the Sun's Rossby number is
not well constrained observationally. Thus, we are biased by those numerical
models that produce dynamos with desirable properties. For example, global-scale
dynamo simulations with low Rossby numbers, on the order ${\rm Ro} \sim 10^{-2}$,
can generate cycling dynamo solutions \citep[e.g.,][]{Ghizaru:2010,Brown:2011,
Racine:2011, Kaepylae:2012, Fan:2014} with equatorward migration of magnetic
field \citep[e.g.,][]{Kaepylae:2013}. When the Rossby number has such a low
value, the spatial scale associated with the thermal Rossby waves is typically
rather small, corresponding to an azimuthal order of $m\sim 100$ \citep{Featherstone:2016}.
Unfortunately, in actual solar observations, such short wavelength waves would
be easily confused with and masked by supergranulation.

All of these difficulties have so far made the search for peaks in the power
spectra of the observed velocity field an unfruitful method to detect thermal
Rossby waves. Instead, identifying thermal Rossby waves may require careful
long-duration averages of the correlations between flow components that are
indicative of the thermal Rossby wave's horizontal wavefunction. Such correlations
may have already been detected years ago by \cite{Schou:2003} and \cite{Gizon:2003}
when they found wave-like properties in the supergranulation signal.

Even if the conjecture that we proposed in Section~\ref{subsec:location_waveguide}
is correct, and the lower half of the convection zone can indeed form a waveguide for
long wavelength thermal Rossby waves, those wave modes may not be visible at the
surface. Assessing the visibility of the gravito-inertial modes would require
knowing how the waves are excited and damped, thus, allowing estimates of the mode
amplitude to be made. Further, since the cavity is confined to the deeper layers of the
convection zone, the wavefunction at the surface (where it can be measured) will
have undergone many e-folding decay lengths. A similar problem exists for the
Sun's $g$ modes. The stellar physics community is certain that gravity modes
must exist in the Sun's radiative interior, but we have not convincingly detected
the action of such modes at the solar surface.

\begin{figure*}
	\epsscale{1.0}
	\plottwo{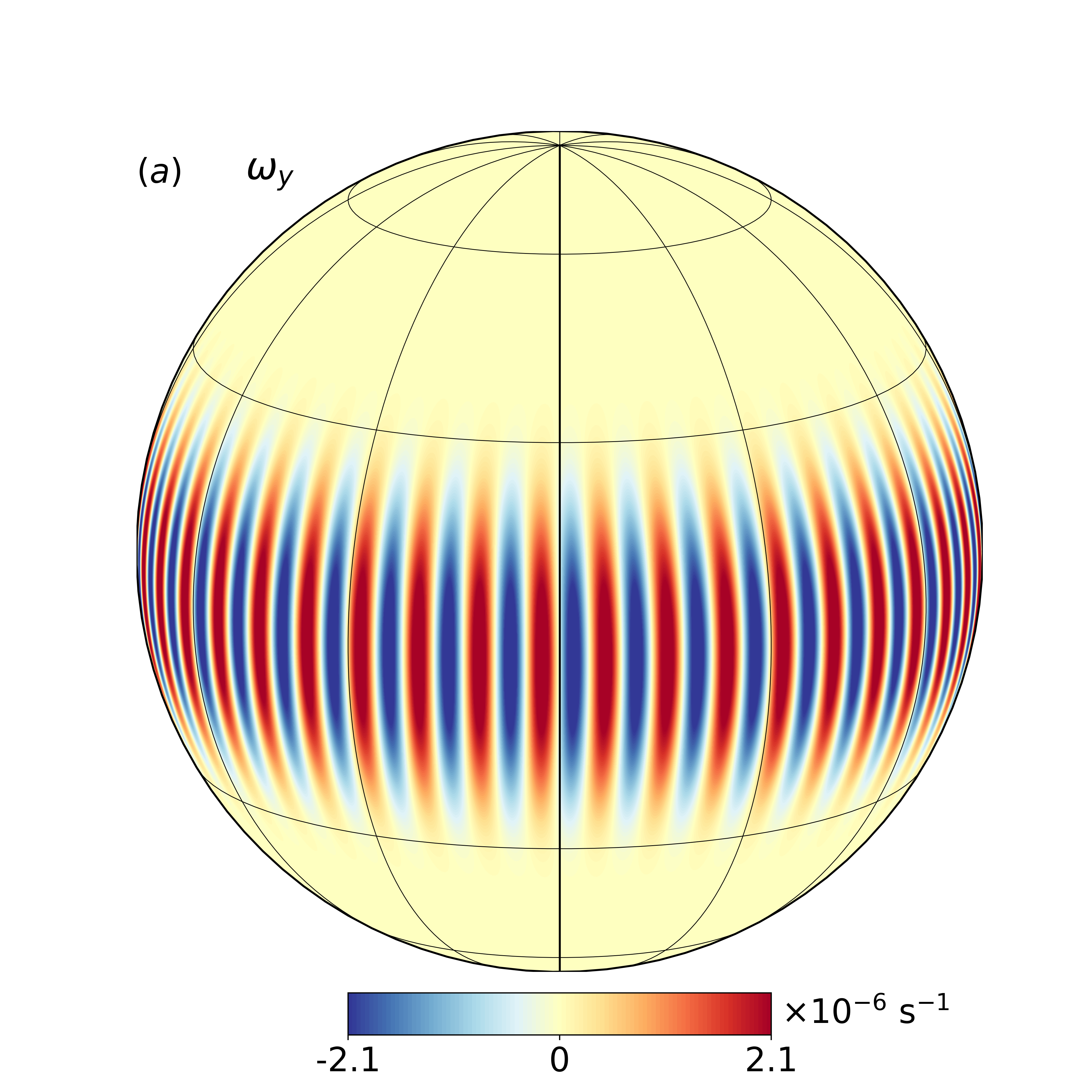}{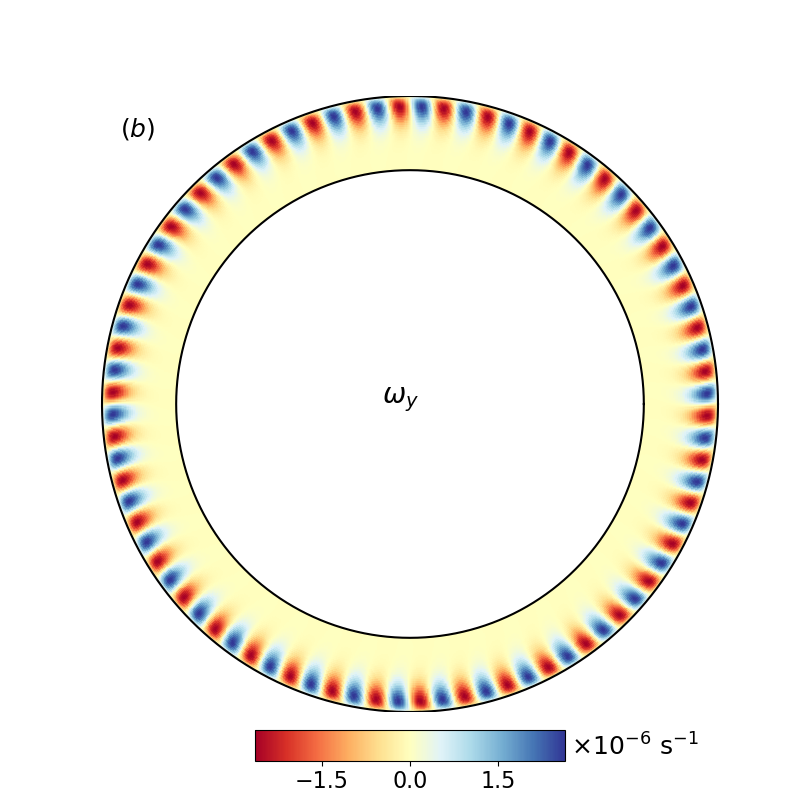}

	\caption{\small An example of a thermal Rossby wave that appears prominently
at convective onset in a 3D numerical simulation of convection in a spherical shell,
specifically Model \#68 from \cite{Hindman:2020a}. ($a$) The axial vorticity
($\omega_y = \bvec{\omega}\cdot\unitv{\Omega}$) in orthographic projection shown on a
spherical surface that lies just below the outer boundary of the computation domain.
Red tones indicate anticyclonic vortices and blue tones cyclonic flow. This particular
simulation produces a wave with 42 complete wavelengths girding the equator ($m=42$).
Each zonal wavelength consists of two counter-spinning rolls whose axes are aligned
with the rotation vector. ($b$) The axial component of the vorticity illustrated on
the equatorial plane as viewed from above the north pole. Red tones indicate anticyclonic
(clockwise) vortices and blue tones correspond to those that are cyclonic (counter-clockwise).
From these images, it is clear that the thermal Rossby wave is trapped radially and
latitudinally; the wave resides in a cavity that is confined to low latitudes and to
radii that lie very close to the outer surface. The numerical simulations that
generated the flow data illustrated by these images was generated by the Rayleigh
convection code \citep{Featherstone:2021, Featherstone:2016, Matsui:2016}.
	\label{fig:thermal_Rossby}}
\end{figure*}

\begin{figure*}
	\epsscale{1.0}
	\plotone{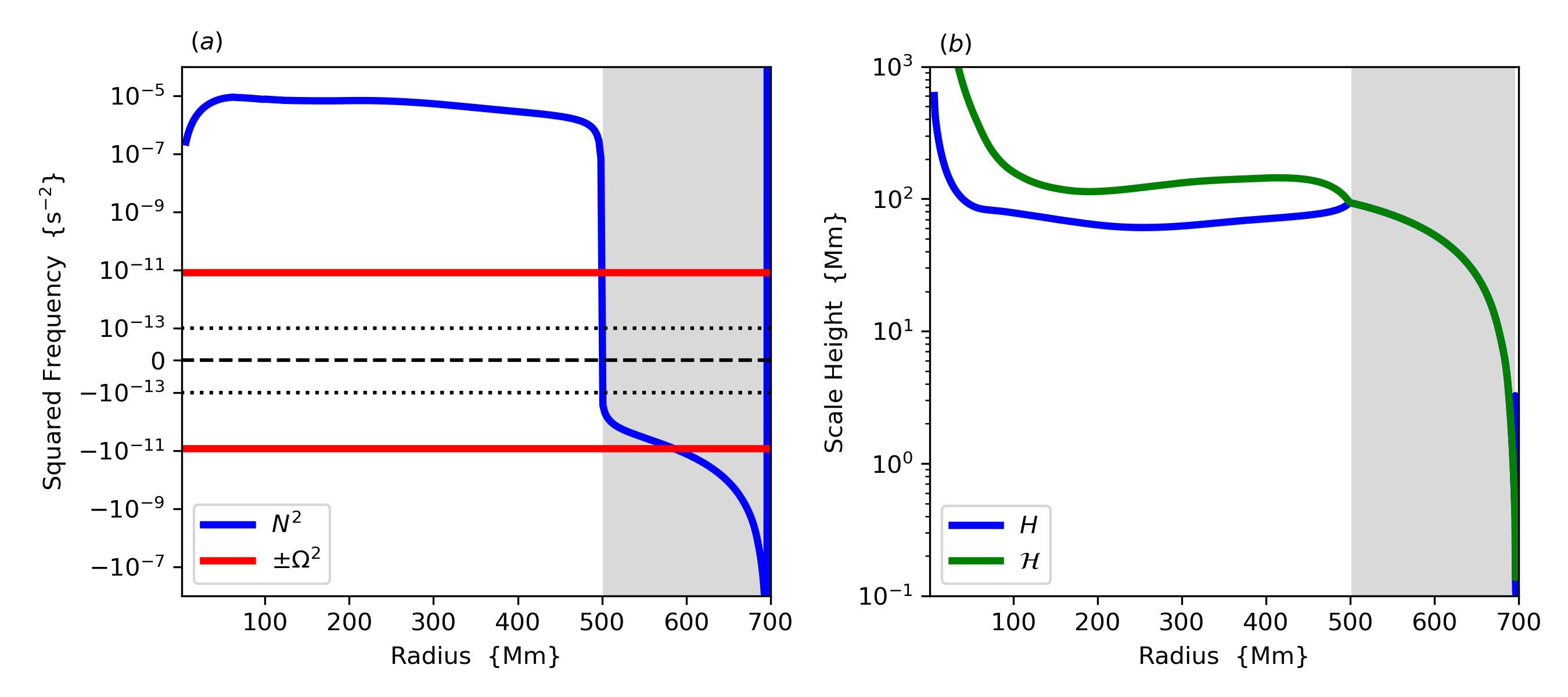}

	\caption{\small Radial atmospheric profiles for the Sun, as specified by
Model S from \cite{Christensen-Dalsgaard:1996}. ($a$) The square of the buoyancy
frequency $N^2$ is illustrated with the blue curve as a function of radius. The
ordinate axes is logarithmically scaled in both the positive and negative values,
with a region of linear scaling in the middle that extends between the dotted lines.
The Sun's convection zone coincides with the region where the square of the buoyancy
frequency is negative and is indicated by the gray shaded region. The two red horizontal
lines mark the square of the Carrington rotation rate, $\pm\Omega\sun^2$, where
$\Omega=\Omega_\sun = 2.87\times 10^{-6}$ s$^{-1}$. The entire radiative interior
is in the regime of slow rotation, $N\gg\Omega$. ($b$) The density scale height $H$
(blue curve) and the scale height $\Hc$ (green curve) shown as functions of radius
in the Sun. The two scale heights are nearly identical in the convection zone and
have roughly the same magnitude throughout much of the radiative interior.
	\label{fig:freq_ratio}}
\end{figure*}

\begin{figure*}
	\epsscale{1.0}
	\plotone{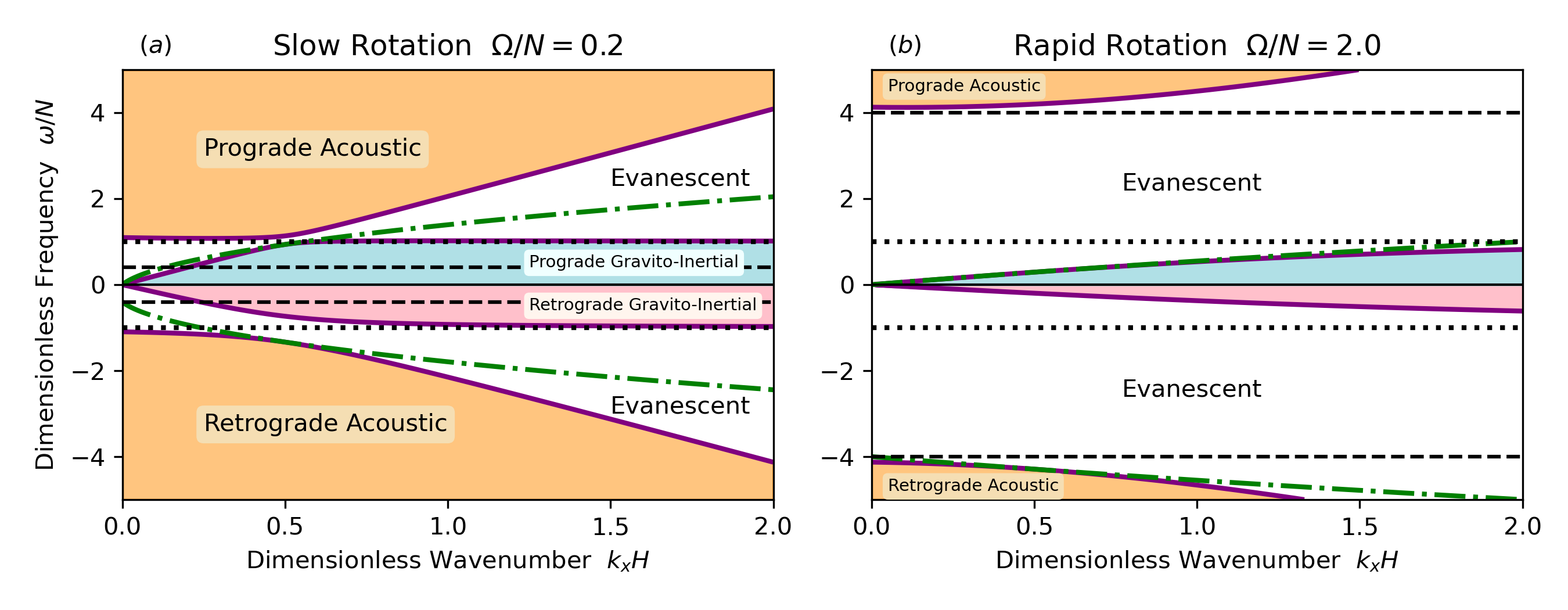}

	\caption{\small Propagation diagrams for an isothermal atmosphere in the
limits of ($a$) slow rotation and ($b$) rapid rotation. For each limit, the ratio
of the buoyancy frequency to the rotation rate is indicated at the top of the panel.
Those frequencies and wavenumbers that correspond to radially propagating waves
appear in the shaded regions. The prograde and retrograde branches of the acoustic
waves appear in orange. The light-blue region indicates gravito-inertial
waves that propagate zonally in the prograde direction and the pink region contains
the retrograde gravito-inertial waves. The horizontal lines indicate the characteristic
frequencies of the atmosphere. The dotted lines mark the positive and negative
values of the buoyancy frequency, $\omega = \pm N$ and the dashed lines bound the
inertial frequency range, $\omega = \pm 2 \Omega$. At low wavenumbers the low-frequency
waves are gravito-inertial waves that have distinctly anisotropic branches. At high
wavenumbers, these waves become nearly isotropic internal gravity waves. The two
dot-dashed green curves indicate the $f$ modes for the rotating star.
	\label{fig:isothermal_propagation}}
\end{figure*}

\begin{figure*}
	\epsscale{0.5}
	\plotone{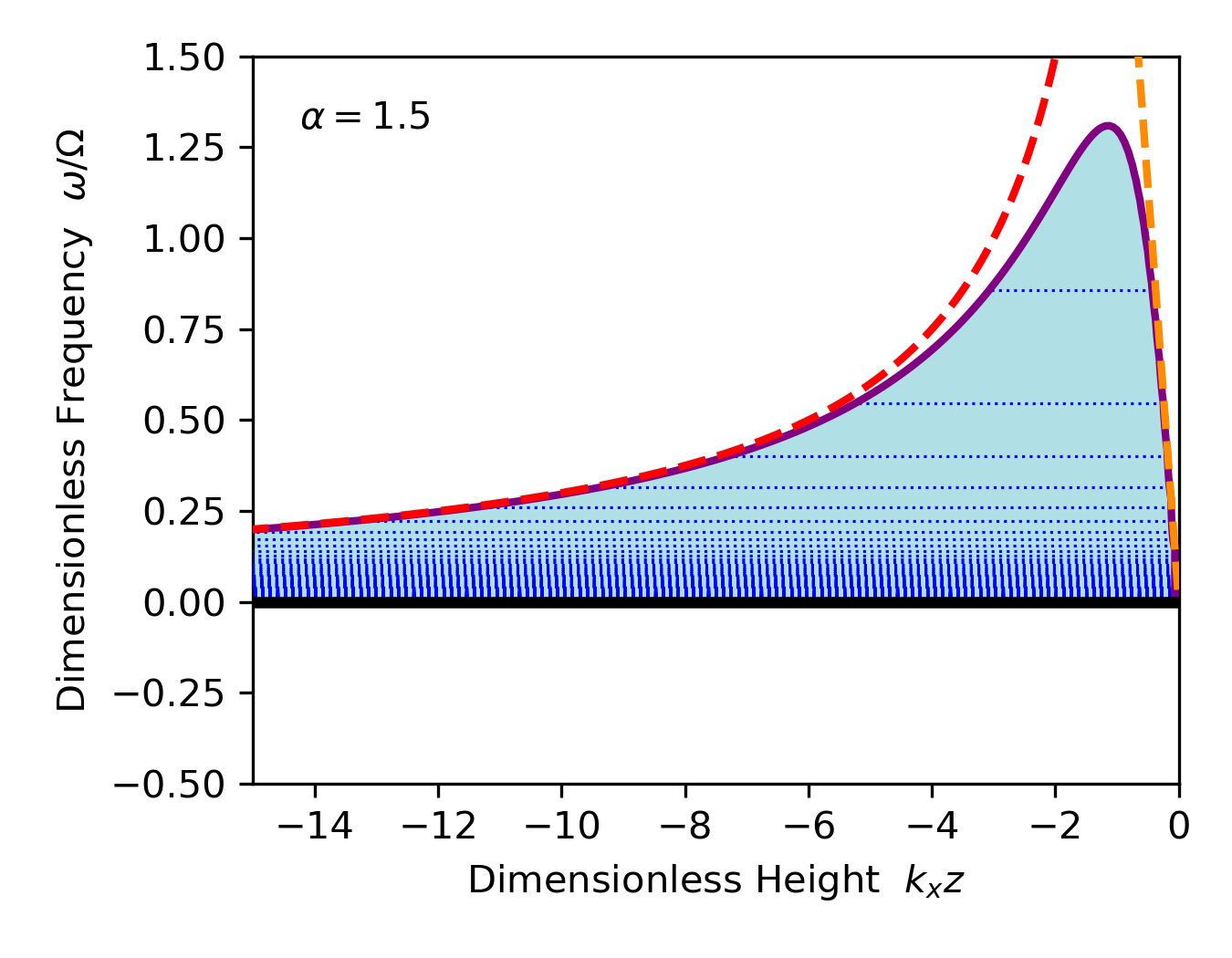}

	\caption{\small Propagation diagram for thermal Rossby waves in an atmosphere
that is neutrally stable to convective overturning. At each height in the atmosphere,
the frequencies of waves that are vertically propagating are shaded pale blue. All
such waves have positive frequencies and propagate longitudinally in a prograde direction
(in the rotating reference frame). The solid purple curve indicates the upper frequency
bound for radial propagation. If one draws a horizontal line at a specific frequency,
the two points where this line crosses the upper bound denote the turning points
and the blue shaded region in between corresponds to the inertial-wave cavity. The
red and orange dashed lines are approximations to the upper frequency bound that
are valid deep in the atmosphere (red) and near the upper surface (orange)---see
the text, Equation~\eqnref{eqn:bounding_frequency}. The thin, horizontal, blue,
dotted lines mark the eigenfrequencies of the thermal Rossby modes. 
	\label{fig:polytropic_propagation}}
\end{figure*}

\begin{figure*}
	\epsscale{1.0}
	\plotone{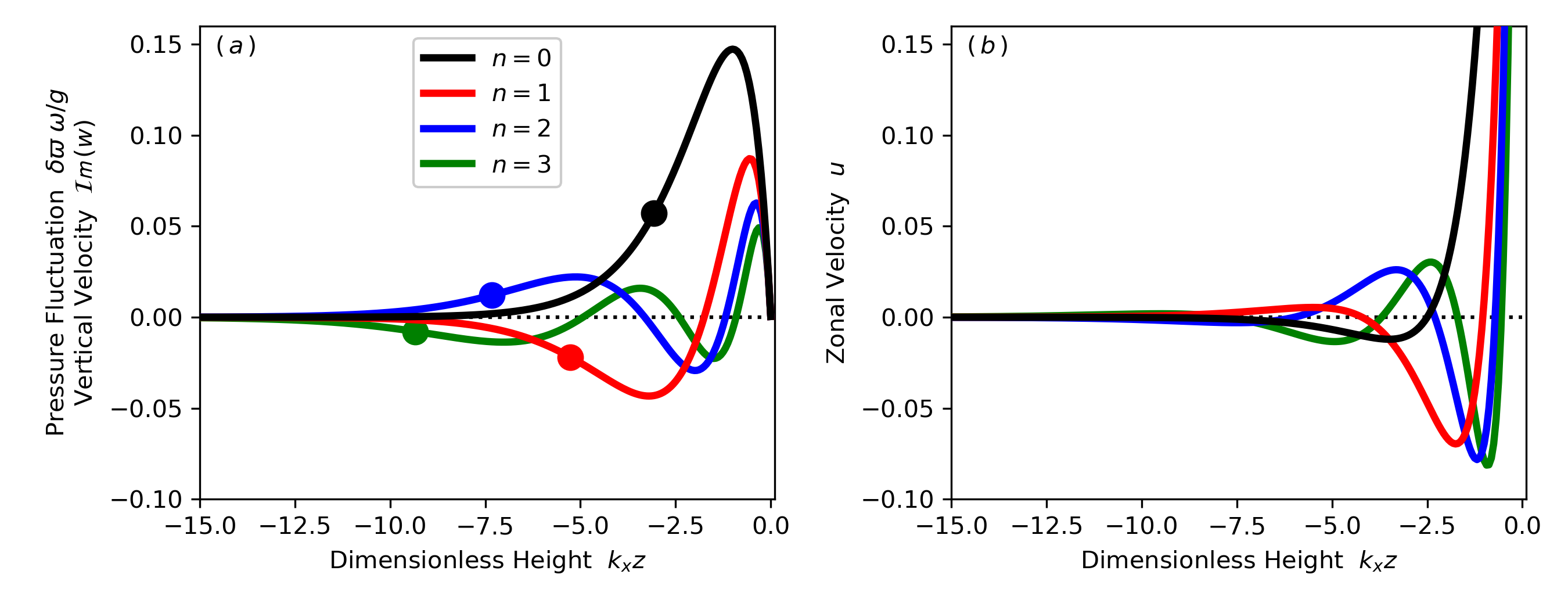}

	\caption{\small Radial eigenfunctions for a semi-infinite neutrally-stable
polytropic atmosphere. The two panels correspond to ($a$) the reduced Lagrangian
pressure fluctuation $\dP$ and ($b$) the zonal velocity $u$. For modes like these
with low frequency, the left-hand panel also indicates the imaginary part of the
vertical velocity $w$. The first four radial modes with radial orders $n = 0$, 1,
2, and 3 are indicated by the color of the curve (black, red, blue, and green,
respectively). The colored dots that appear on the eigenfunction curves indicate
the location of the lower turning point of the wave cavity. The upper turning
points are all very close to the upper boundary, $z=0$, and are omitted for clarity.
Note, the eigenfunctions only depend on the zonal wavenumber through a self-similar
scaling of the vertical coordinate. Thus, these eigenfunctions are universal and
apply to all modes of the relevant radial order, independent of the zonal wavelength.
	\label{fig:polytropic_eigenfunctions}}
\end{figure*}

\begin{figure*}
	\epsscale{0.5}
	\plotone{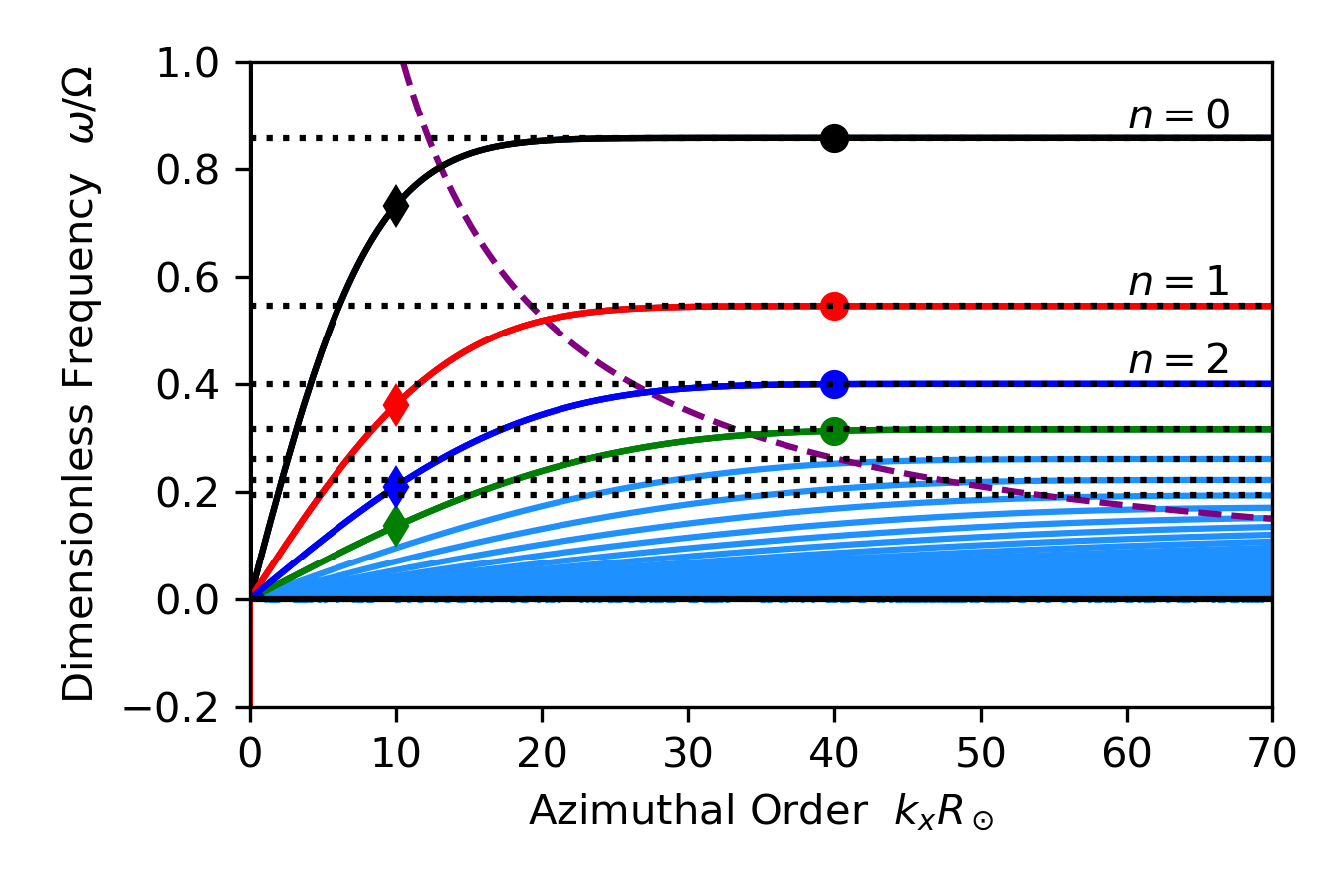}

	\caption{\small Eigenfrequencies for a neutrally stable layer of finite depth
shown as a function of the zonal wavenumber $k_x$. The upper boundary of the layer
is located at the origin, and the bottom of the layer is 200 Mm deep mimicking the
Sun's convection zone. At the upper boundary we require that the solutions remain
finite and on the lower boundary we impose a condition of impenetrability. The 
eigenfrequencies of the first four radial modes are shown with the black ($n=0$),
red ($n=1$), blue ($n=2$), and green ($n=3$) curves. Higher order overtones ($n>3$)
are drawn in pale blue. At very short zonal wavelengths (high $k_x$) the eigenfunctions
do not sense the presence of the lower boundary. Hence, they have the same frequencies
as the semi-infinite layer. These asymptotic frequencies are indicated with the dotted
horizontal lines. As the wavenumber decreases (and the zonal wavelength grows), the
lower edge of the wave cavity approaches the bottom of the domain and eventually
crosses it. The frequency for which this crossing occurs is indicated by the dashed
purple curve. For wavenumbers less than this threshold, the eigenfunctions (and
eigenfrequencies)  are strongly influenced by the lower boundary condition and the
frequency decreases linearly towards zero frequency at zero wavenumber. Along the
first four dispersion curves are marked two modes. The diamonds indicate modes with
wavenumber $k_x R_\sun = 10$ and the circles correspond to modes with $k_x R_\sun = 40$.
The eigenfunctions for these modes are illustrated in Figure~\ref{fig:finitedepth_eigfuncs}.
	\label{fig:finitedepth_eigfreqs}}
\end{figure*}

\begin{figure*}
	\epsscale{1.0}
	\plotone{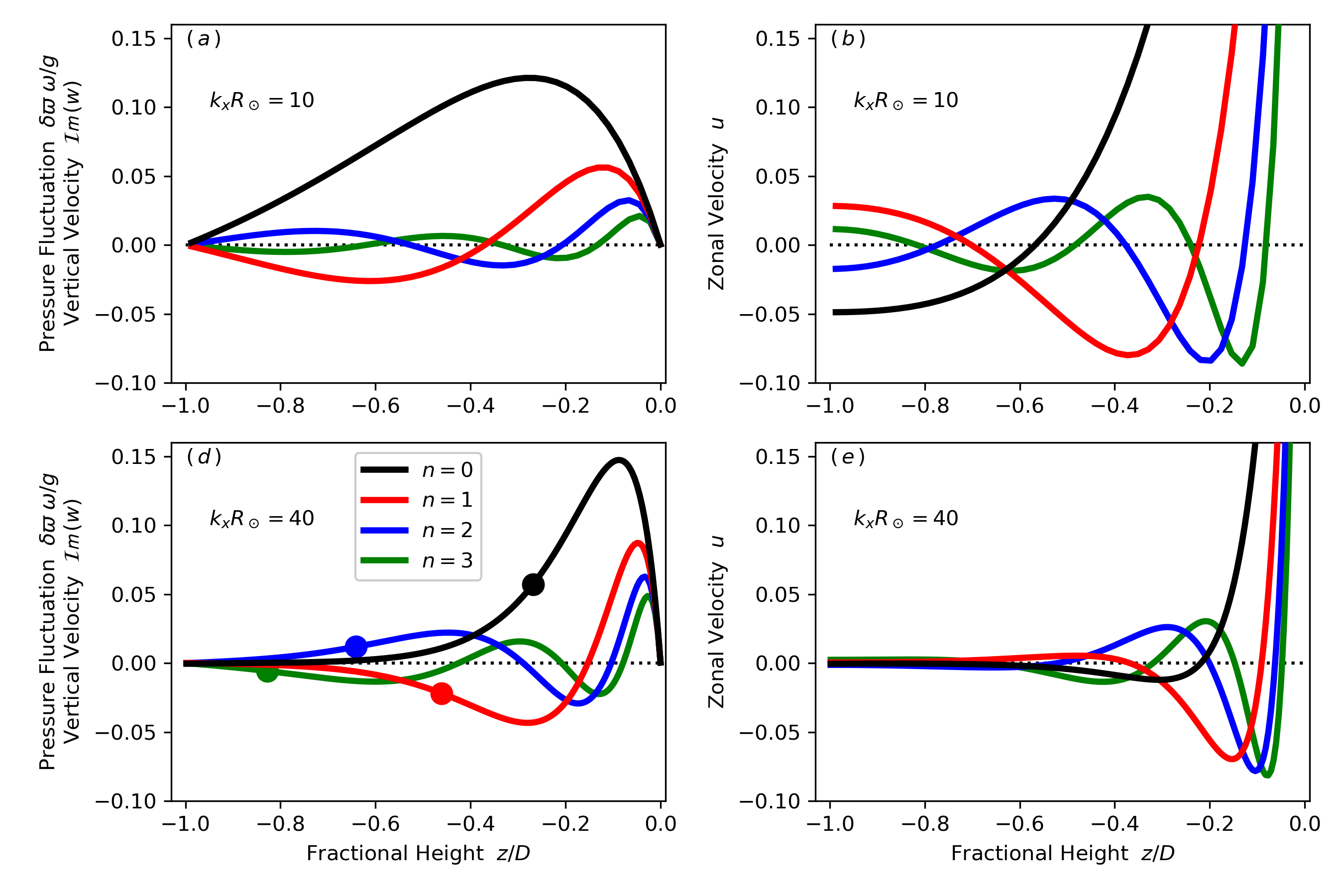}

	\caption{\small Eigenfunctions for the modes indicated by the diamonds
(top panels) and circles (bottom panels) that appear in Figure~\ref{fig:finitedepth_eigfreqs}.
The upper panels correspond to modes with a low zonal wavenumber given by $k_x R_\sun = 10$,
and the lower panels to a larger wavenumber $k_x R_\sun = 40$. The left-hand panels,
($a$) and ($c$), show the vertical variation of the reduced Lagrangian pressure
fluctuation, $\dP(z)$, and simultaneously the imaginary part of the vertical velocity
$w(z)$. The right-hand panels, ($b$) and ($d$), present the zonal velocity $u(z)$.
The eigenfunctions for first four radial orders, $n=0$, 1, 2, and 3, appear as the
black, red, blue, and green curves, respectively. Using the same color scheme, the
lower turning point for each mode is indicated by a large circular dot. For the modes
that appear in the upper panels, the lower turning point is below the lower boundary,
and thus unillustrated outside the solution domain.
	\label{fig:finitedepth_eigfuncs}}
\end{figure*}

\begin{figure*}
	\epsscale{0.5}
	\plotone{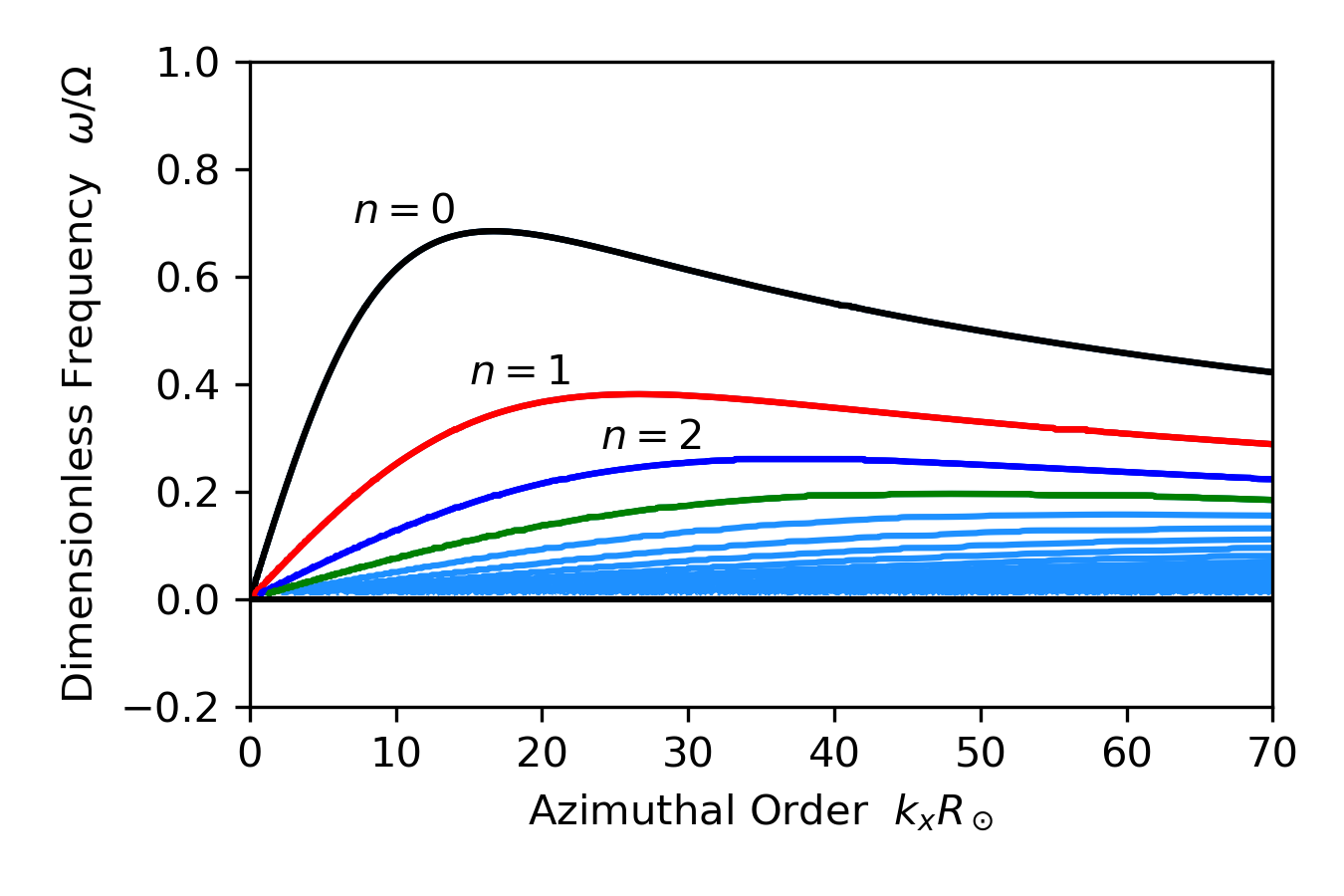}

	\caption{\small Eigenfrequencies as a function of zonal wavenumber $k_x$
for a fully submerged polytropic layer with lower and upper boundaries at $z=-200$ Mm
and $z=-30$ Mm, respectively. A condition of impenetrability is applied at both
boundaries. The colors have the same meaning as in Figure~\ref{fig:finitedepth_eigfreqs}.
For low wavenumbers, the eigenfrequencies increase linearly and this behavior is caused
by the lower turning point being below the lower boundary of the convection zone. The
upper boundary condition exerts its influence most strongly at large wavenumbers. For
these modes, the upper turning point lies above the upper boundary of the domain,
and the wave cavity is effectively truncated by the upper boundary condition. A smaller
wave cavity requires smaller frequencies in order to squeeze the same number of
vertical wavelengths within the cavity.  Hence, at sufficiently large wavenumber,
the eigenfrequencies decrease monotonically with increasing wavenumber as the wave
cavity becomes increasingly truncated.
	\label{fig:finitedomain_eigfreqs}}
\end{figure*}

\begin{figure*}
	\epsscale{1.0}
	\plotone{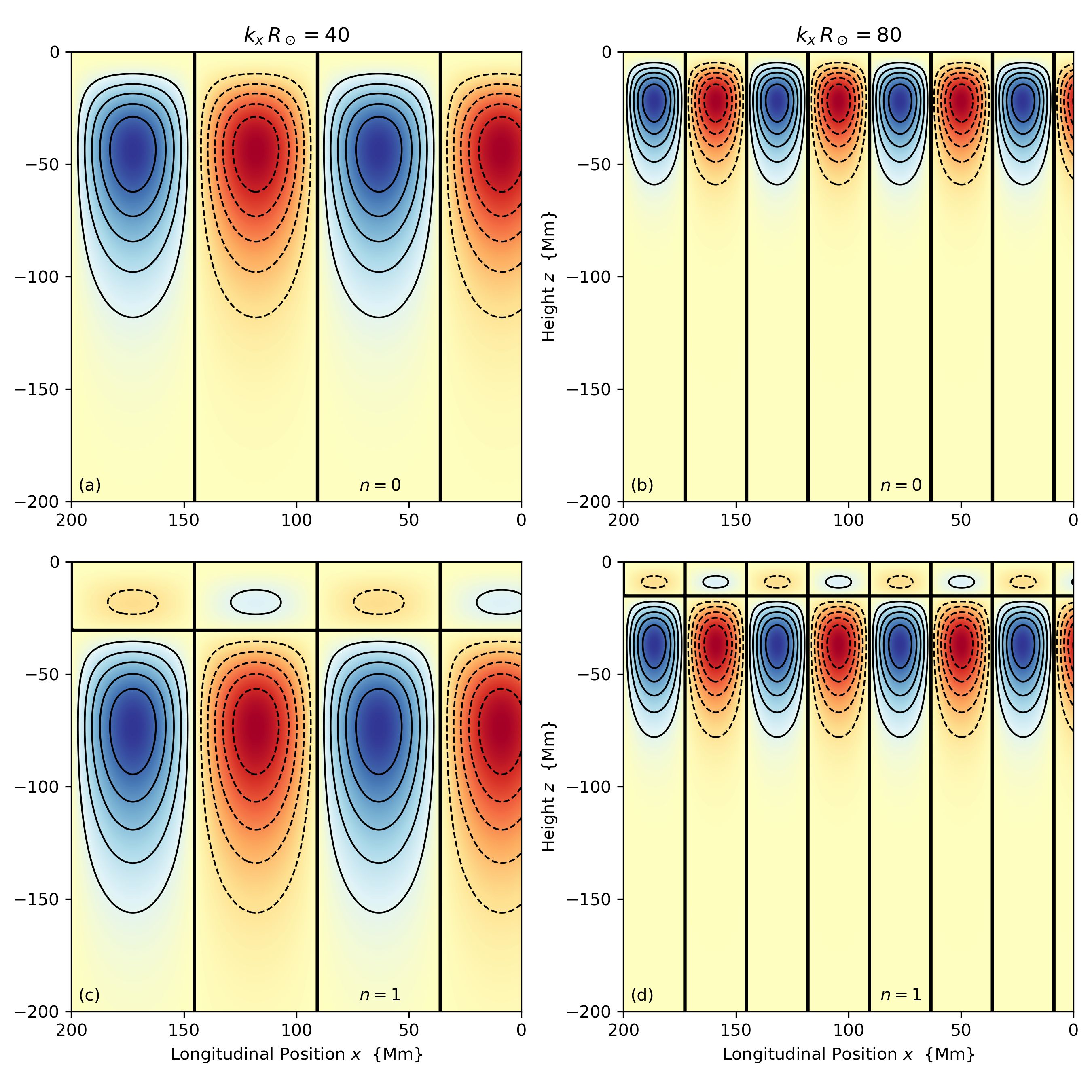}

	\caption{\small The stream function for the mass flux for four eigenmodes
within a semi-infinite polytropic atmosphere. Red tones with dashed black isocontours
indicate negative values of the stream function and blue tones and solid isocontours
positive values. The flow field is being viewed on a section of the equatorial plane as
viewed from above the north pole. Motion in this plane is clockwise (anticyclonic)
along the isocontours when the stream function is negative, and counter-clockwise
(cyclonic) along the contours for positive values. The upper row of panels illustrates
the vortex structure for two fundamental radial modes $n=0$ with different zonal wavenumbers,
as indicated at the top of each panel. The lower panels correspond to the first radial
overtone $n=1$. The fundamental mode consists of single vortex or roll in radius, and
two counter-rotation rolls within a zonal wavelength. Each successive overtone stacks
another roll in depth. The shape of the rolls does not change as the zonal wavenumber
increases. Instead the spatial scale in all dimensions changes, maintaining the aspect
ratio of the vortices.  
	\label{fig:polytropic_cells}}
\end{figure*}

The very sensitivity of the thermal Rossby waves to the superadiabatic gradient
which leads to all of the nuisances in the observability of those modes, would
make the thermal Rossby waves, if detected, an excellent seismic diagnostic of
the superadiabaticity in the convection zone. Since, the thermal Rossby waves tend
to be confined near the equator---see Figure~\ref{fig:thermal_Rossby}, they would
be most sensitive to the gradient at low latitudes. It might be possible to assess
weak latitudinal variations in the buoyancy frequency if the information from
thermal Rossby waves were to be combined with observations of the high-latitude
inertial oscillations that were recently discovered by \cite{Gizon:2021}. As
discussed in \cite{Gizon:2021} the high-latitude modes also possess sensitivity
to the superadiabaticity, but in the polar regions.


\acknowledgments

We would like to thank Jon Aurnou, Maria Camisassa, Keith Julien, and Lydia Korre
for useful discussions and providing early feedback that influenced the development
of this manuscript. This work was supported by NASA through grants 80NSSC17K0008,
80NSSC18K1125, 80NSSC19K0267, and 80NSSC20K0193. R.J. would like to acknowledge the
support of MSRC (SoMaS), University of Sheffield (UK) and is grateful to Science and
Technology Facilities Council (STFC) grant ST/V000977/1.




\end{document}